\documentclass[prd,aps,nofootinbib,superscriptaddress,floatfix,preprintnumbers,twocolumn]{revtex4}

\usepackage[colorlinks, citecolor=blue,anchorcolor=blue,menucolor=blue, linkcolor=blue,filecolor=blue,runcolor=blue,urlcolor=blue,frenchlinks=true]{hyperref}
\usepackage{amsmath,amsfonts,amssymb,mathtools,cases,slashed,bm}
\usepackage{epsfig, graphicx}
\usepackage{color}
\usepackage[dvipsnames]{xcolor}
\usepackage{tabularx}
\usepackage{physics}
\usepackage{hhline}
\usepackage{mathrsfs}
\usepackage[normalem]{ulem}
\usepackage{multirow}
\usepackage{float}
\usepackage{appendix}
\usepackage{orcidlink}
\usepackage[normalem]{ulem}

\begin{document}

\title{ Systematic Study of the Self-Renormalized Nucleon Gluon PDF  in Large-Momentum Effective Theory}

\author{Alex NieMiera \orcidlink{0009-0007-0553-2590}}
\email{niemiera@msu.edu}
\affiliation{Department of Physics and Astronomy, Michigan State University, East Lansing, Michigan 48824, USA}
\affiliation{Department of Computational Mathematics,
  Science and Engineering, Michigan State University, East Lansing, Michigan 48824, USA}

\author{William Good \orcidlink{0000-0001-8428-1003}}
\email{goodwil9@msu.edu}
\affiliation{Department of Physics and Astronomy, Michigan State University, East Lansing, Michigan 48824, USA}
\affiliation{Department of Computational Mathematics,
  Science and Engineering, Michigan State University, East Lansing, Michigan 48824, USA}

\author{Huey-Wen Lin \orcidlink{0000-0001-6281-944X}}
\affiliation{Department of Physics and Astronomy, Michigan State University, East Lansing, Michigan 48824, USA}

\author{Fei Yao \orcidlink{0000-0002-7227-0996}}
\affiliation{Physics Department, Brookhaven National Laboratory, Upton, New York 11973, USA}

\begin{abstract}
We present a systematic study of the nucleon gluon parton distribution function (PDF) using the self-renormalized large-momentum effective theory (LaMET) approach in lattice QCD.
This work extends previous gluon-PDF extractions by performing a detailed analysis of key systematic effects, including gauge-link smearing, lattice spacing, pion mass, and nucleon boost momentum.
The self-renormalization framework mitigates ultraviolet divergences associated with Wilson-line self-energy and renormalon contributions by combining lattice matrix elements with perturbative short-distance information, thereby preserving the correct infrared structure.
Calculations are performed on $N_f=2+1+1$ HISQ ensembles generated by the MILC Collaboration at three lattice spacings and two pion masses, with boosted nucleon states reaching momenta up to 2.2~GeV.
We determine renormalization factors from zero-momentum matrix elements and apply hybrid renormalization to suppress discretization artifacts.
After extrapolating large-separation behavior and performing Fourier transforms, we reconstruct quasi-PDFs and match them to lightcone PDFs using next-to-leading order Wilson coefficients.
Our results demonstrate that smearing and lattice-spacing effects are under control, and pion-mass and lattice-spacing dependence is mild relative to the current $O(10^6)$ statistics; however, momentum dependence remains a significant source of uncertainty.
Future work including even larger boost momenta will be essential to reduce systematics in lattice determinations of the gluon PDF and to advance toward precision QCD phenomenology at the LHC and the future Electron-Ion Collider.
\end{abstract}
\preprint{MSUHEP-25-025}
\date{\today}
\maketitle

\section{Introduction}
\label{sec:intro}

Parton distribution functions (PDFs) are nonperturbative quantities that encode the momentum distributions of quarks and gluons inside hadrons.
They are essential inputs for theoretical predictions of high-energy scattering processes, such as those measured at the Large Hadron Collider (LHC).
In particular, uncertainties in the gluon PDF limit the theoretical precision of key processes, including Higgs-boson production, jet production, and heavy-flavor dynamics, which are central to testing the Standard Model and probing potential new physics.
While the quark structure of the proton is relatively well constrained by deep inelastic scattering (DIS) and Drell-Yan measurements, these processes are only indirectly sensitive to gluons, especially in the intermediate- and large-$x$ regions.
Constraining the gluon PDF in these kinematic domains, therefore, relies on sparser data sets, such as inclusive jet production and top-quark observables~\cite{Hou:2019efy}.
Differences in data cuts and theoretical treatments across global analyses have led to a lack of consensus on the gluon distribution, underscoring the need for improved constraints~\cite{Amoroso:2022eow, Achenbach:2023pba}.
Ongoing and future measurements at the LHC, as well as the upcoming Electron-Ion Collider (EIC), are expected to significantly reduce gluon PDF uncertainties and improve our understanding of proton structure~\cite{Accardi:2012qut,AbdulKhalek:2021gbh,Burkert:2022hjz}.

In contrast to global PDF extractions, which rely on phenomenological fits to a wide range of experimental cross sections, lattice QCD provides a systematically improvable, first-principles approach to accessing the nonperturbative structure of hadrons.
Over the past decade, significant progress has been made in computing $x$-dependent parton distributions on the lattice through methods such as large-momentum effective theory (LaMET)~\cite{Ji:2013dva} and the pseudo-PDF framework~\cite{Balitsky:2019krf}.
These approaches have enabled determinations of momentum-fractions-dependent shapes for isovector nucleon, valence-pion quark PDFs and generalized parton distributions, as reviewed in Refs.~\cite{Ji:2020ect,Constantinou:2020hdm,Lin:2023kxn,Lin:2025hka}.
However, lattice calculations of gluon PDFs remain considerably more challenging.
The number of available studies is much smaller, and the associated uncertainties are typically larger than for quark distributions.
Most existing work has employed the pseudo-PDF method with simplified functional parameterizations for the $x$ dependence of the hadron gluon distribution~\cite{Fan:2020cpa,Fan:2022kcb,HadStruc:2021wmh,Delmar:2023agv,HadStruc:2022yaw,Good:2023ecp,Salas-Chavira:2021wui,NieMiera:2025inn}.
Only recently have the first extractions of the nucleon gluon PDF using LaMET become feasible, enabled by improved control of the long-distance behavior of the relevant matrix elements~\cite{Good:2024iur,Good:2025daz,NieMiera:2025mwj}.

In this work, we extend our recent determination of the nucleon gluon parton distribution using the self-renormalized LaMET approach~\cite{NieMiera:2025mwj} by performing a detailed study of several key systematic effects, including gauge-link smearing, lattice spacing, pion mass, and nucleon boost momentum.
This represents an important step toward achieving full systematic control in lattice QCD extractions of the gluon PDF using LaMET.

The self-renormalization procedure~\cite{LatticePartonCollaborationLPC:2021xdx} removes divergences associated with the Wilson-line self-energy, renormalon contributions, and both leading and subleading logarithmic ultraviolet behavior by combining lattice matrix elements with perturbative short-distance information.
In doing so, it mitigates lattice-spacing artifacts while preserving the correct infrared structure of the matrix elements, avoiding distortions that can arise in purely nonperturbative renormalization schemes.
Until now, this framework has been applied primarily to quark bilinear observables~\cite{LatticeParton:2022xsd,LatticeParton:2022zqc,Holligan:2023rex,Holligan:2024wpv}, and our study marks its first systematic application to the gluon sector.
We show that the self-renormalization method provides robust control over smearing- and lattice-spacing--related uncertainties.
Additionally, pion mass effects are comparatively mild, while the dependence on the nucleon boost momentum remains one of the leading sources of systematic uncertainty, requiring further refinement in future calculations.

The rest of this paper is organized as follows: in Sec.~\ref{sec:self-renorm} we give the lattice details of our calculation, describe self-renormalization of the gluon operator, and show our self-renormalized zero momentum matrix elements.
In Sec.~\ref{sec:numerical_results}, we explore smearing and lattice spacing effects in our self-renormalized matrix elements, demonstrate the large-$\nu$ extrapolations, show the effects of matching, and explore various systematic effects on the PDFs.
We conclude and make remarks on future work in Sec.~\ref{sec:conclusions}.

\section{Lattice Details and Self-Renormalization Factors}
\label{sec:self-renorm}

\subsection{Lattice Details}
For this study, we analyze high-statistics measurements on gauge-field ensembles generated by the MILC Collaboration~\cite{MILC:2013znn}, corresponding to lattice spacings of approximately $a \approx 0.15$, $0.12$, and $0.09$~fm.
The ensembles contain $N_f = 2+1+1$ flavors of highly improved staggered quarks (HISQ) in the sea sector~\cite{Follana:2007rc}, with valence pion masses of $M_{\pi} \approx 310$ and $690$~MeV.
In the valence sector, we employ clover-improved Wilson fermions, with the quark masses tuned to match the light and strange quark masses in the HISQ sea.
This mixed-action setup follows the strategy extensively used by the PNDME Collaboration in previous precision studies of nucleon structure over multiple lattice spacings and quark masses~\cite{Park:2025rxi,Jang:2023zts,Mondal:2020cmt,Jang:2019jkn,Gupta:2018lvp,Lin:2018obj,Gupta:2018qil,Bhattacharya:2015wna,Bhattacharya:2015esa,Bhattacharya:2013ehc}.
No evidence of exceptional configurations was observed in either the two-point or three-point correlation functions on any of the ensembles considered in this work.

To determine the self-renormalization factors, we generated approximately $O(10^5)$ two-point correlation functions for the $\eta_s$ meson at zero momentum.
For the nucleon matrix elements relevant to the PDF extractions, we computed $O(10^6)$ two-point correlators at both the light and heavy pion masses, employing Gaussian momentum smearing to improve overlap with boosted states~\cite{Bali:2016lva}.
The $\eta_s$ correlators were constructed using the $\overline{s}\gamma_t\gamma_5 s$ interpolating operator, while the nucleon interpolating operator was taken as $\epsilon_{abc}(d_a^T C \gamma_5 u_b)\mathcal{P}+ u_c$, where $C$ denotes the charge-conjugation matrix and $\mathcal{P}+$ projects onto positive-parity states.
Although alternative high-momentum--optimized operator constructions have recently been proposed in Ref.~\cite{Zhang:2025hyo}, much of the correlator data used in this analysis was produced prior to their introduction.
Additionally, the improvement associated with these newer operators scales inversely with the hadron mass, and thus is expected to be less pronounced for the heavier-than-physical pion masses considered here.
Details of the lattice ensembles and simulation parameters are provided in Table~\ref{tab:lattice_details}.
For convenience, we refer to the nucleon states obtained at the lighter and heavier pion masses as $N_l$ and $N_s$, respectively.
Following the methodology used in the MSULat program~\cite{Fan:2020cpa, Salas-Chavira:2021wui, Fan:2022kcb, Good:2023ecp, Good:2024iur, NieMiera:2025inn,  Good:2025daz}, two- and three-point correlation functions are analyzed using simultaneous two-state fits, with the fit windows optimized to suppress excited-state contamination.

\begin{table*}[t]
    \centering
    \renewcommand{\arraystretch}{1.3}
    \begin{tabular}{| c | c | c | c |}
    \hline
        Ensemble & a$09$m$310$ & a$12$m$310$ & a$15$m$310$ \\ \hline
        $a$ (fm) & $0.0888(8)$ &  $0.1207(11)$ & $0.1510(20)$ \\ \hline
        $L^3 \cross T$ & $32^3 \cross 96$ & $24^3 \cross 64$ & $16^3 \cross 48$ \\ \hline
        $M_\pi^\text{val}$ (GeV) & $0.313(1)$ & $0.309(1)$ & $0.319(3)$ \\ \hline
        $M_{\eta_s}^\text{val}$ (GeV) & $0.698(7)$ & $0.6841(6)$ & $0.687(1)$ \\ \hline
        $P_z$ (GeV) & \{0, 1.31, 1.75, 2.18\} & \{0, 1.28, 1.71, 2.14\} & \{0, 1.54, 2.05, 2.57\} \\ \hline
        $N_\text{cfg}$ & $1026$ & $1013$ & $900$ \\ \hline
        $N_\text{meas}^\text{2pt}$ ($P_z = 0$) & $196,992$ & $64,832$ & $129,600$ \\ \hline
        $N_\text{meas}^\text{2pt}$ ($P_z \neq 0$) & $1,378,944$ & $1,555,968$ & $950,400$ \\ \hline
    \end{tabular}
    \caption{For each HISQ ensemble generated by the MILC Collaboration and used in this analysis, we list the lattice spacing $a$, lattice volume $L^3 \times T$, the valence pion mass $M_\pi^\text{val}$, the valence $\eta_s$ mass $M_{\eta_s}^\text{val}$, and the values of the hadron momenta $P_z$ employed. We also provide the number of gauge configurations $N_\text{cfg}$ analyzed, together with the number of two-point correlator measurements performed at zero momentum, $N_\text{meas}^\text{2pt}(P_z = 0)$, and at nonzero momentum, $N_\text{meas}^\text{2pt}(P_z \neq 0)$.}
    \label{tab:lattice_details}
\end{table*}

Previous lattice studies have demonstrated that reliable signals for gluon correlators typically require the use of gauge-link smearing~\cite{Fan:2020cpa, Fan:2022kcb, HadStruc:2021wmh, Salas-Chavira:2021wui,  HadStruc:2022yaw,Delmar:2023agv, Good:2023ecp,Hackett:2023rif,Hackett:2023nkr, Good:2024iur, NieMiera:2025inn, Good:2025daz}.
In contrast, the analytical formulation of self-renormalization is derived for unsmeared Wilson lines, although it has been shown that the method remains valid as long as the smearing radius is sufficiently small compared with the inverse boost momentum~\cite{LatticePartonLPC:2021gpi}.
Because a fully established framework for incorporating smearing into self-renormalization is not yet available, we take this opportunity to assess the impact of different smearing prescriptions by performing three complementary analyses, each employing a different choice of Wilson flow~\cite{Luscher:2010iy} in the construction of the gluon operators.

Wilson flow is characterized by a flow time $\mathcal{T}_\text{W} = t_\text{W} a^2$, where $t_\text{W}$ is a dimensionless parameter and $a$ is the lattice spacing.
Our first analysis uses a uniform relative smearing choice with $t_\text{W}=3$ for all ensembles.
This corresponds to physical flow times ranging from $\mathcal{T}_\text{W}\approx 0.024\text{ fm}^2$ on the finest lattice to $\mathcal{T}_\text{W}\approx 0.068\text{ fm}^2$ on the coarsest lattice, resulting in approximately constant smearing in lattice units. We refer to this approach as the fixed relative smearing method.
To contrast with this, we also examine two fixed physical smearing schemes, in which we select flow times to maintain approximately constant $\mathcal{T}_\text{W}$ across lattice spacings.
Specifically, we employ $t_{\text{W}} = \{1,2,3\}$ on the coarsest ensemble and $t_{\text{W}} = \{2,3,5\}$ on the finest one, which correspond to physical flow-time ranges of $\mathcal{T}_{\text{W}} \approx 0.023$--$0.029\text{ fm}^2$ and $\mathcal{T}_{\text{W}} \approx 0.040$--$0.046\text{ fm}^2$, respectively.
For convenience, we label these three analyses as W$333$ (fixed relative smearing), W$123$, and W$235$ (fixed physical smearing), and use this notation throughout the remainder of the paper.
In the W333 analysis, a fixed relative smearing is used, for which $\mathcal{T}_W = t_W a^2$ scales with $a^2$ and the smoothing scale $1/\sqrt{8\mathcal{T}_W}$ remains proportional to $1/a$, preserving the standard $1/a$ linear divergence and ensuring the validity of the
self-renormalization formula.
For W123 and W235, we instead employ fixed-physical smearing, where $\mathcal{T}_W$ is approximately constant across lattice spacings;
although the linear divergence no longer scales as $1/a$, the resulting modulation can be absorbed into the renormalon term $m_{0}$ and the nonperturbative function $g(z)$ in our parameterization, leaving the multiplicative state-independent UV structure intact and allowing the same self-renormalization relation to be used.

\subsection{Self-Renormalization Factor Determination}

In the LaMET framework, we begin with the bare spatially separated matrix elements,
\begin{equation}
    \tilde{h}^{B}(z, P_z, 1/a) = \langle P \,|\, O(z) \,|\, P \rangle,
\end{equation}
where $|P\rangle$ denotes a hadron state boosted to momentum $P_z$, and $O(z)$ is a gauge-invariant nonlocal gluon operator with separation $z$.
Although several multiplicatively renormalizable gluon operators can, in principle, be used to access the gluon PDF~\cite{Zhang:2018diq,Balitsky:2019krf}, we find that the operator introduced in Ref.~\cite{Balitsky:2019krf} provides the most reliable signal for extracting renormalized matrix elements.
This operator is defined as
\begin{equation} \label{eq:gluon-operator}
    O(z) = F^{t i}(z)\, U(z,0)\, F^{t}_{\ i}(0)
           - F^{i j}(z)\, U(z,0)\, F_{i j}(0),
\end{equation}
where the indices $i,j$ denote the transverse directions $\{x,y\}$, and $U(z,0)$ is a Wilson line that ensures gauge invariance.
The gluon field-strength tensor is given by
\begin{equation}
    F^{\mu\nu}_{a} = \partial^\mu A^\nu_{a}
                     - \partial^\nu A^\mu_{a}
                     - g\, f_{abc}\, A^\mu_{b} A^\nu_{c}.
\end{equation}

To remove the ultraviolet (UV) divergences present in the bare quasi-correlation function $\tilde{h}^{B}(z, P_{z}, 1/a)$, one introduces a multiplicative renormalization factor~\cite{Zhang:2018diq}:
\begin{equation} \label{eq:multiplicative-renormalization}
    \tilde{h}^{B}(z, P_{z}, 1/a)
    =
    Z_L(a) \, e^{-\delta m(a)|z|} \, \tilde{h}^{R}(z, P_{z}),
\end{equation}
where $\tilde{h}^{R}$ denotes the renormalized matrix element.
The function $Z_L(a)$ cancels the logarithmic divergences that do not depend on the Wilson-line length $z$, while the exponential factor containing $\delta m(a)$ removes the Wilson-line self-energy divergence that grows linearly with $z$.
Fully nonperturbative renormalization prescriptions~\cite{Chen:2016fxx,Izubuchi:2018srq,Alexandrou:2017huk,Radyushkin:2018cvn,Braun:2018brg,Li:2018tpe}, however, typically suffer from undesirable nonperturbative artifacts introduced by the renormalization factor itself, which can alter the infrared behavior of the quasi-correlation function.
The hybrid renormalization approach~\cite{Ji:2020brr} mitigates this issue by treating short- and long-distance contributions separately.
In this scheme, the renormalized correlation is defined as
\begin{multline} \label{eq:renormalized-quasi-LF}
    \tilde{h}^{R}(z, P_{z})
    =
    \frac{\tilde{h}^{B}(z, P_{z}, 1/a)}{\tilde{h}^{B}(z, P_{z}=0, 1/a)} \, \theta(z_{s} - |z|) \\
    + \; T_{s} \, \frac{\tilde{h}^{B}(z, P_{z}, 1/a)}{Z_{R}(z, 1/a)} \, \theta(|z| - z_{s}),
\end{multline}
where $z_{s}$ specifies the separation between the short- and long-distance regions, and $T_{s}$ is a matching factor chosen to ensure smooth behavior at $|z| = z_{s}$.
For $|z| < z_{s}$, the ratio method~\cite{Radyushkin:2018cvn} is employed to remove the $z$-dependent UV divergences associated with the Wilson line.
For $|z| > z_{s}$, we instead apply a self-renormalization procedure in which $Z_{R}(z, 1/a)$ is obtained from a global fit of zero-momentum matrix elements evaluated at multiple lattice spacings.

To carry out self-renormalization, we compute the renormalization factors using matrix elements evaluated in the hadron rest frame.
The bare matrix element at zero momentum, $\tilde{h}^B (z, P_z=0, 1/a)$, is connected to the corresponding renormalized matrix element $\tilde{h}^R (z, P_z=0)$ via a multiplicative renormalization factor:
\begin{equation} \label{eq:multaplicative_ZR}
    \tilde{h}^B (z, P_z=0, 1/a) = Z_R (z, 1/a) \, \tilde{h}^R (z, P_z=0).
\end{equation}
The factor $Z_R (z, 1/a)$ is computed at one-loop order within the operator-product-expansion framework and depends explicitly on the operator under consideration.
Its functional form can be expressed as
\begin{multline} \label{eq:renormalization-factor}
    Z_R (z, 1/a) = \exp \Biggl[ \frac{k z}{a \ln(a \Lambda_{\text{QCD}})} + m_0 z + f(z) a^2 \\
    + \frac{5 C_A}{3 b_0} \ln \left( \frac{\ln[1/(a \Lambda_{\text{QCD}})]}{\ln[\mu / \Lambda_{\text{QCD}}]} \right)
    + \ln \left( 1 + \frac{d}{\ln(a \Lambda_{\text{QCD}})} \right) \Biggr].
\end{multline}
In this expression, the first term represents a linear divergence in the lattice spacing, the $m_0 z$ term encodes finite mass contributions arising from renormalization ambiguities, and $f(z) a^2$ parametrizes discretization effects associated with the lattice action.
The final two terms account for the resummation of leading and sub-leading logarithmic divergences.

In the $\overline{\text{MS}}$ scheme, the next-to-leading order (NLO) Wilson coefficient for the gluon is given by~\cite{Good:2025daz}:
\begin{equation} \label{eq:gluon-wilson-coeff}
    C_{0, \text{NLO}}(z, \mu)
    =
    1 + \frac{\alpha_s(\mu) C_A}{4 \pi} \left[ \frac{5}{3} \ln \left( \frac{z^2 \mu^2}{4 e^{-2 \gamma_E}} \right) + 3 \right],
\end{equation}
where $\gamma_E$ denotes the Euler-Mascheroni constant.
To extract the parameters in Eq.~\ref{eq:renormalization-factor}, we adopt the methodology of Ref.~\cite{Chen:2024rgi} and perform a global fit using the ansatz
\begin{equation}\label{eq:parameterization}
\begin{split}
    \ln \tilde{h}^B(z, P_z = 0, 1/a)
    &=
    \ln \left[ Z_R(z, 1/a)\right]  \\ &
    + \begin{cases}
       \ln[C_{0,\text{NLO}}(z,\mu)] , & z_0 \le z \le z_1, \\[1mm]
        g(z) - m_0 z, & z > z_1,
    \end{cases}
\end{split}
\end{equation}
treating $k$, $\Lambda_\text{QCD}$, $f(z)$, $d$, $m_0$, and $g(z)$ as free parameters.
The fit is applied to interpolated zero-momentum $\eta_s$ matrix elements at lattice spacings $a \approx 0.09$, $0.12$, and $0.15$~fm.
We use $\eta_s$ as the external state since the renormalization factors should be state-independent.

Our previous LaMET study~\cite{Good:2025daz} verified that both the linear divergence and the renormalon ambiguity are statistically consistent across different hadron states for a single lattice spacing.
Here, we consider three smearing setups, W$123$, W$234$, and W$333$, with $z_0 = 0.05$~fm and $z_1 = 0.36$~fm, fixing the renormalization scale at $\mu = 2$~GeV.
A more detailed investigation of the scale dependence is deferred to future work.
The fitted parameters and the corresponding $\chi^2/\text{dof}$ for each global fit are summarized in Table~\ref{tab:fitted_params}.
All $\chi^2/\text{dof}$ values are close to unity, indicating that each fit is equally reliable.

\begin{table}[h!]
    \centering
    \renewcommand{\arraystretch}{1.3}
    \begin{tabular}{| c | c | c | c |}
    \hline
        Parameter & W$123$ & W$235$ & W$333$ \\
        \hline
        $k$ & $1.02(52)$ & $0.660(27)$ & $0.633(26)$ \\
        \hline
        $d$ & $-2.22(13)\times10^{-13}$ & $-1.23(53)\times 10^{-14}$ & $-0.0008(78)$ \\
        \hline
        $\Lambda_\text{QCD}$ (GeV) & $0.133(54)$ & $0.140(52)$ & $0.264(77)$ \\
        \hline
        $m_0$ (GeV) & $0.39(43)$ & $0.099(20)$ & $0.13(25)$ \\
        \hline
        $\chi^2 / \text{dof}$ & $1.04(96)$ & $0.89(71)$ & $0.81(62)$ \\
    \hline
    \end{tabular}
    \caption{
 A summary of the fitted parameters and the corresponding goodness of fit ($\chi^2 / \text{dof}$) from the global fit to Eq.~\ref{eq:parameterization} is also provided.}
    \label{tab:fitted_params}
\end{table}

For the W$333$ case, the physical interpretation of individual parameters should be treated with caution, as quantities such as the linear divergence are expected to depend on the smearing radius.
In contrast, for W$123$ and W$235$, we observe that both the linear divergence $k$ and the renormalon parameter $m_0$ are larger for smaller smearing radii, consistent with the expectation that increased smearing reduces the decay rate of the correlators, effectively lowering the sum of $k$ and $m_0$.
The parameter $d$ is statistically consistent with zero across all three analyses, suggesting limited sensitivity to subleading logarithmic effects at the current statistical precision.
The extracted $\Lambda_\text{QCD}$ values lie within reasonable ranges of 150--300~MeV.
Finally, we note that the fit parameters are strongly correlated, such that large standard deviations do not necessarily indicate poor control over the fits, a feature also highlighted in the original self-renormalization study~\cite{LatticePartonLPC:2021gpi}, which demonstrates extended valleys of acceptable $\chi^2/\text{dof}$ among correlated parameters.

\begin{figure}
\centering
\includegraphics[width=0.45\textwidth]{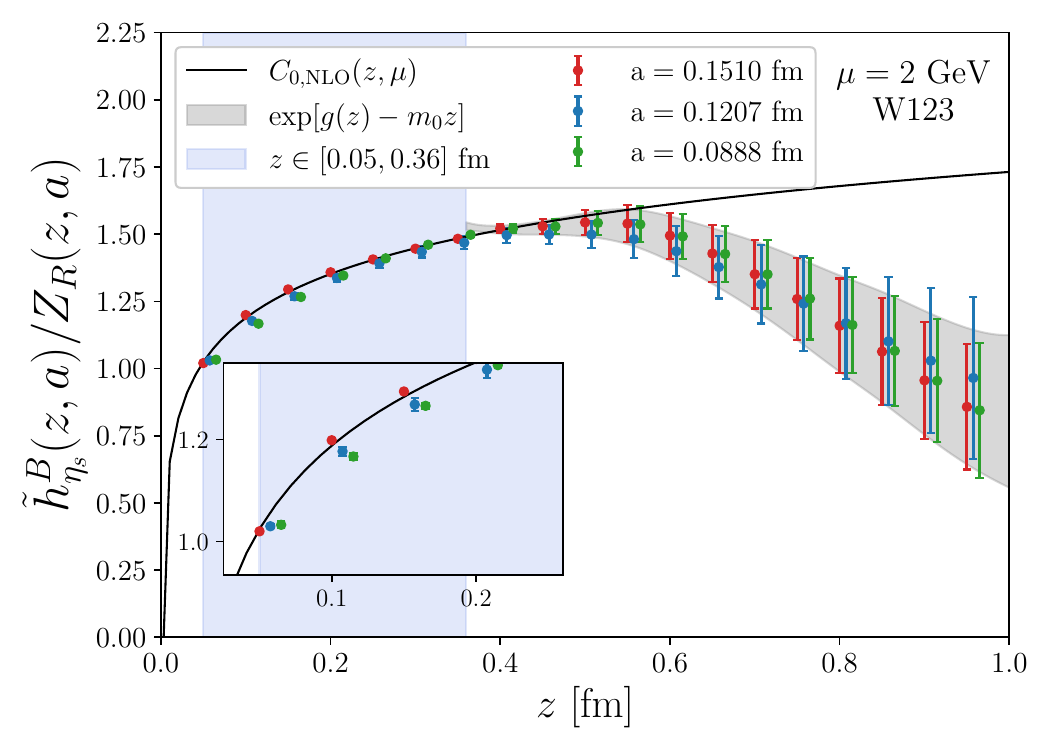}
\includegraphics[width=0.45\textwidth]{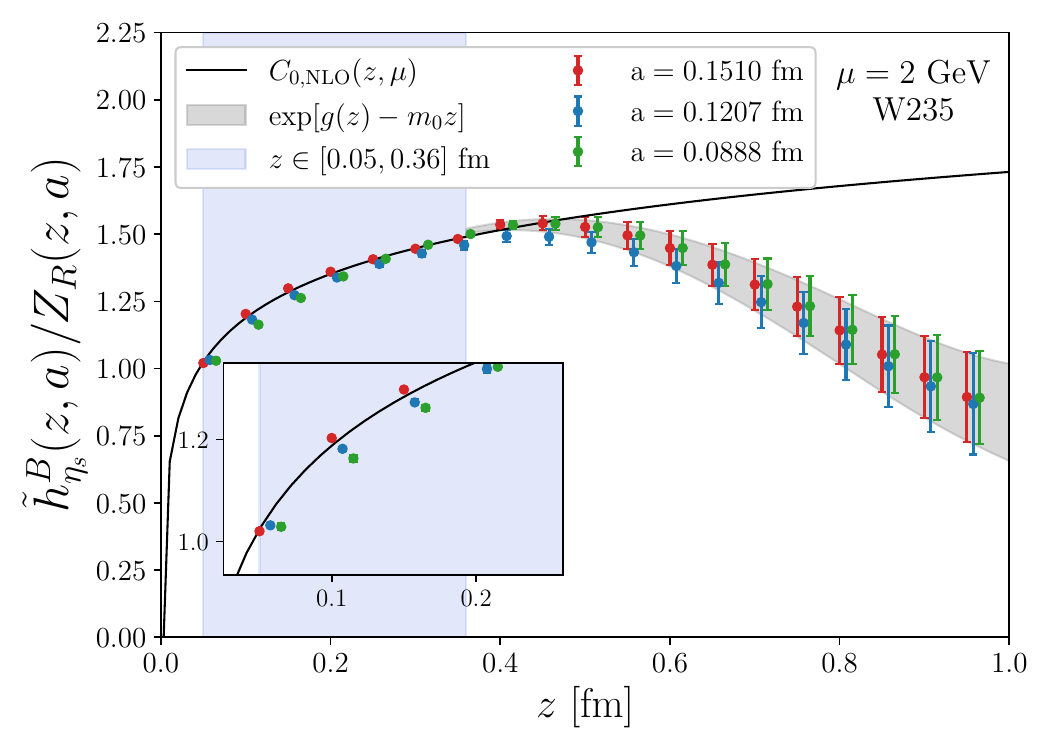}
\includegraphics[width=0.45\textwidth]{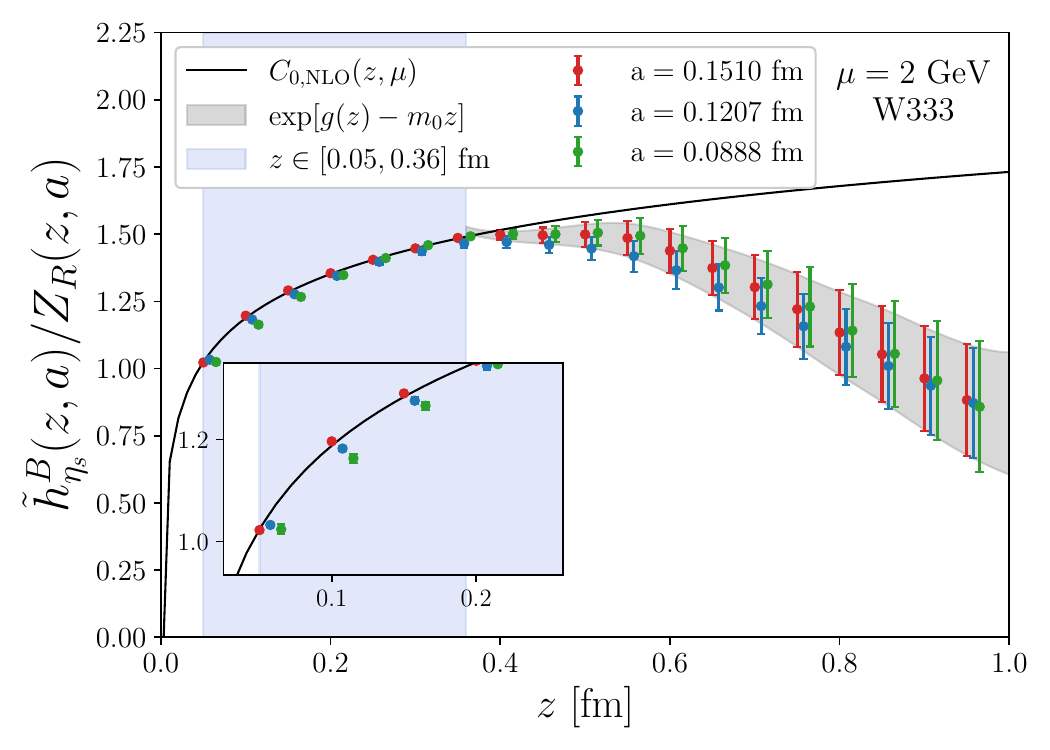}
\caption{
\label{fig:ZR_plots}
The renormalized matrix elements for analyses (top) W$123$,
 (middle) W$235$, and (bottom) W$333$ compared to the perturbative one-loop $\overline{\text{MS}}$ result illustrated by the black curve.}
\end{figure}

We further compare the renormalized zero-momentum values with the perturbative Wilson coefficient given in Eq.~\ref{eq:gluon-wilson-coeff} for the three analyses, as shown in Fig.~\ref{fig:ZR_plots}.
At short distances, the matrix elements exhibit reasonable agreement with the perturbative Wilson coefficients, as expected, while noticeable deviations emerge at larger distances.
All three analyses display similar qualitative trends;
however, the W$123$ analysis appears to follow perturbation theory over a somewhat wider range than the other two cases with greater smearing.
This difference may stem from the relatively larger statistical fluctuations in the latter analyses.
Among the three, the W$235$ analysis yields the smallest uncertainties, likely due to the enhanced smearing applied to the finest lattice spacing.

\section{LaMET Numerical Results}
\label{sec:numerical_results}

\begin{figure}
\centering
\includegraphics[width=0.45\textwidth]{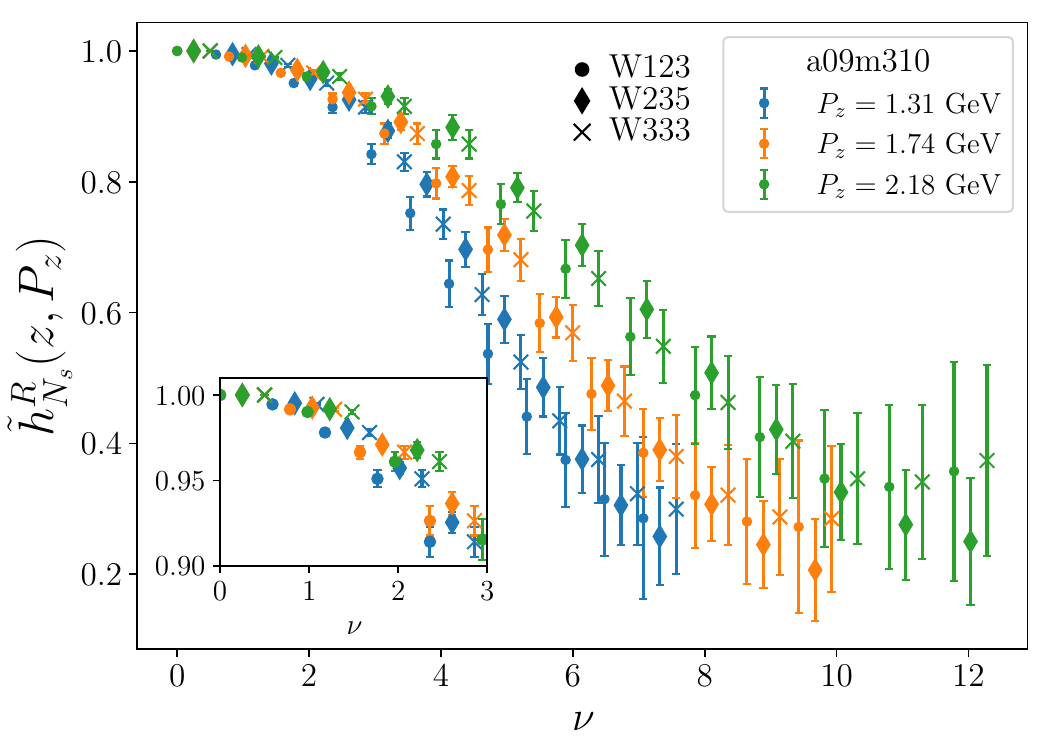}
\includegraphics[width=0.45\textwidth]{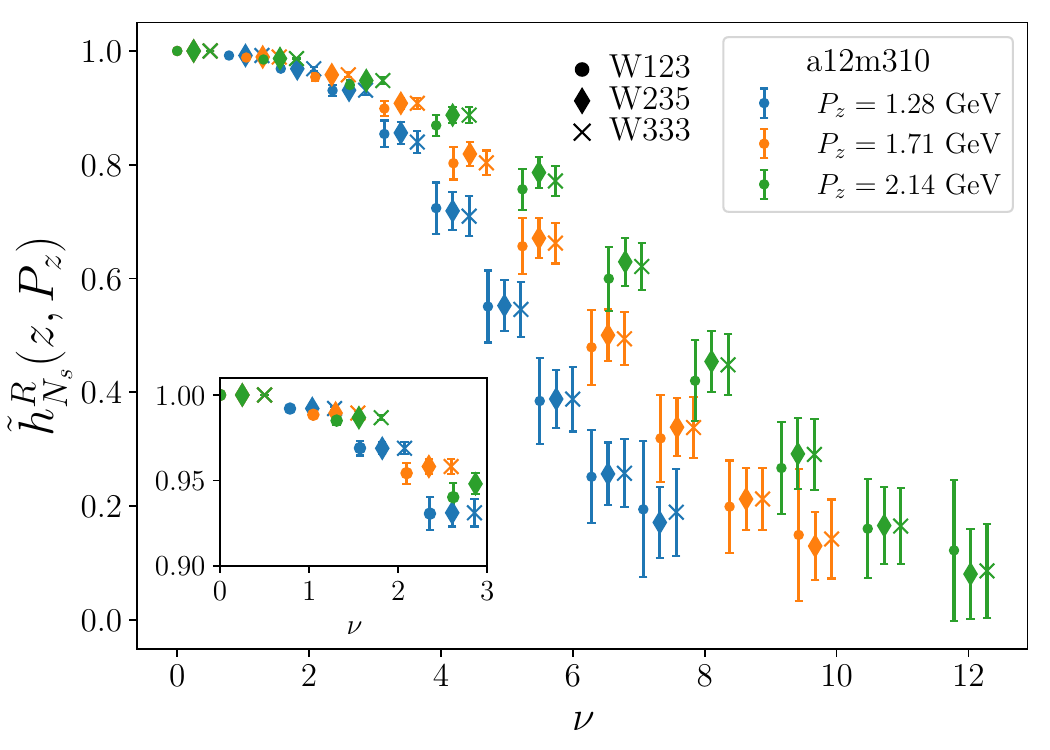}
\includegraphics[width=0.45\textwidth]{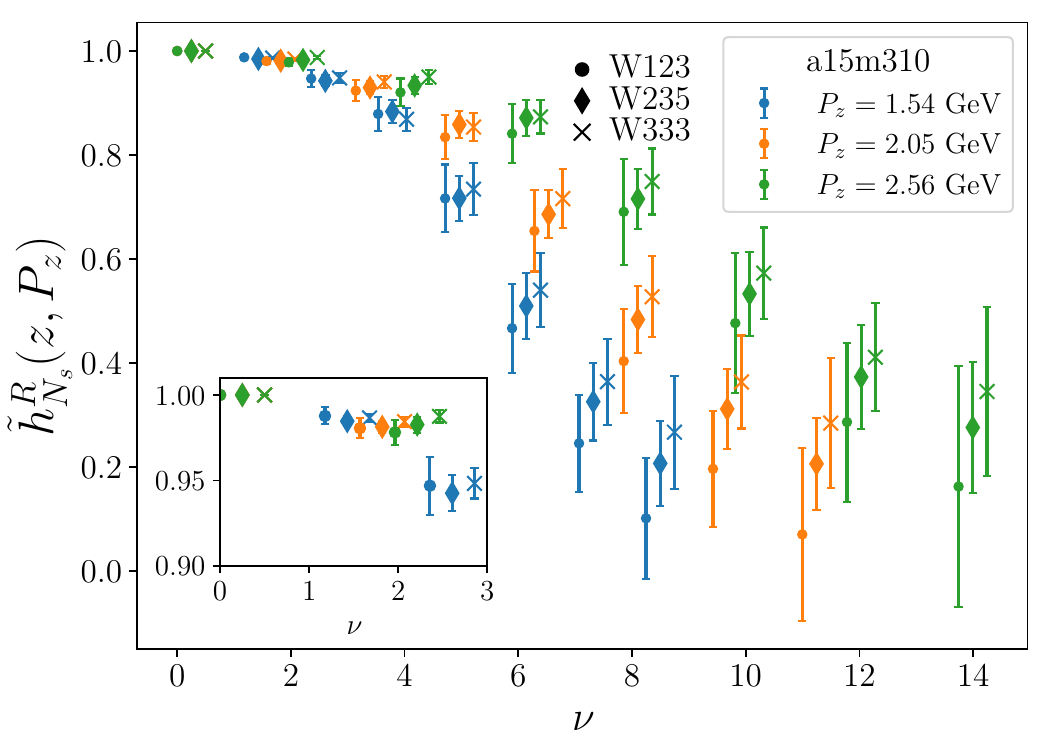}
\caption{
\label{fig:renorm_MEs_ncls}
The renormalized gluon matrix elements at $M_\pi \approx 690$~MeV are compared across the three smearing schemes at three boosted momenta $P_z$ and lattice spacings $a \approx 0.09$~fm (top), $0.12$~fm (middle), and $0.15$~fm (bottom).
}
\end{figure}

With the renormalization factor determined, we examine how the different smearing prescriptions influence the resulting matrix elements.
Fig.~\ref{fig:renorm_MEs_ncls} displays the renormalized results at three nonzero momenta for the ensembles with $M_\pi \approx 690$~MeV.
This heavier pion mass is chosen because it typically yields reduced statistical fluctuations.
The hybrid-renormalization reference scale is fixed to $z_s = 0.36$~fm, as indicated in Eq.~\ref{eq:renormalized-quasi-LF}.
Beginning with the $a \approx 0.09$~fm ensemble (upper panel in Fig.~\ref{fig:renorm_MEs_ncls}), we compare W$123$ (circles) and W$333$ (crosses), which differ by whether the smearing is held fixed in physical or relative units during the self-renormalization procedure, while maintaining the same Wilson-line smearing of $\mathcal{T}_{\text{W}} = 3a^2$ on this ensemble.
The resulting matrix elements show excellent statistical agreement: both central values and uncertainties coincide over all values of $z$ and $P_z$.
The W$235$ analysis (diamonds), which employs a slightly larger total smearing $\mathcal{T}_\text{W} = 5a^2$, exhibits marginally reduced statistical uncertainties, while remaining consistent within one standard deviation of the other two determinations.

\begin{figure*}
\centering
\includegraphics[width=0.45\textwidth]{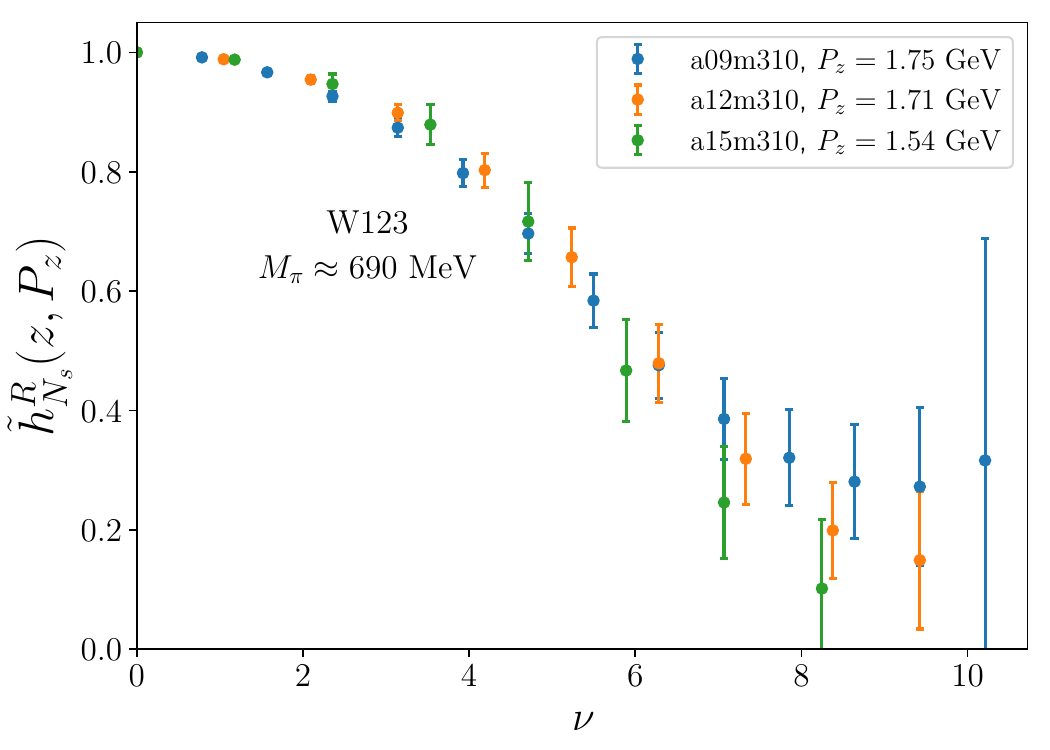}
\includegraphics[width=0.45\textwidth]{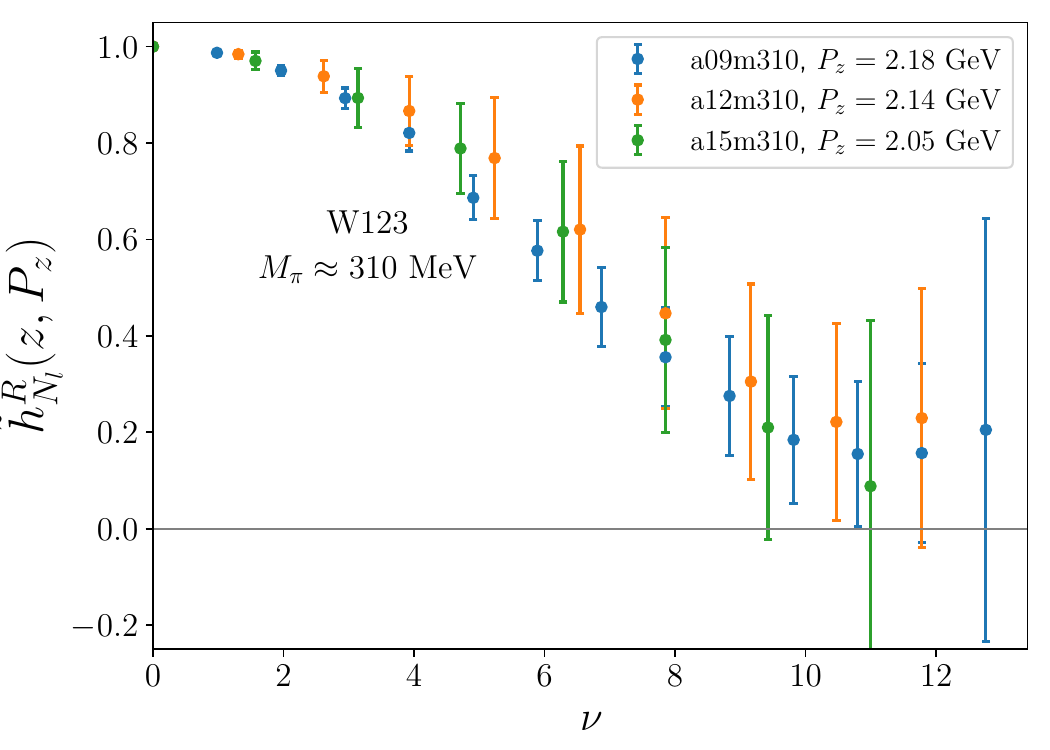}
\includegraphics[width=0.45\textwidth]{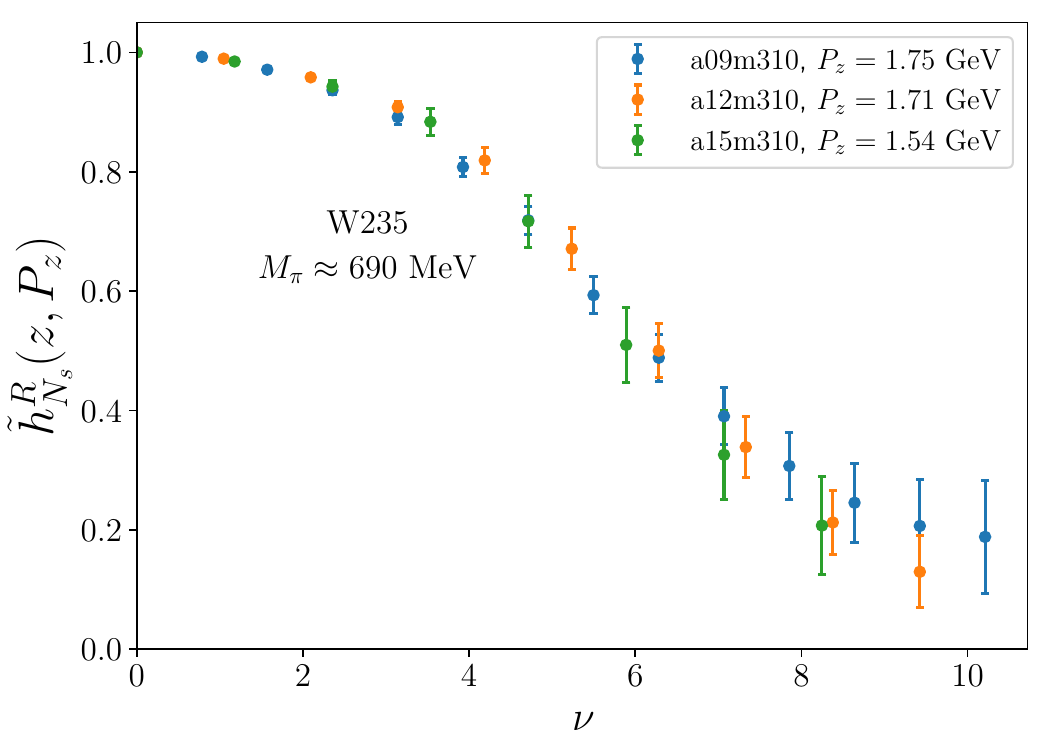}
\includegraphics[width=0.45\textwidth]{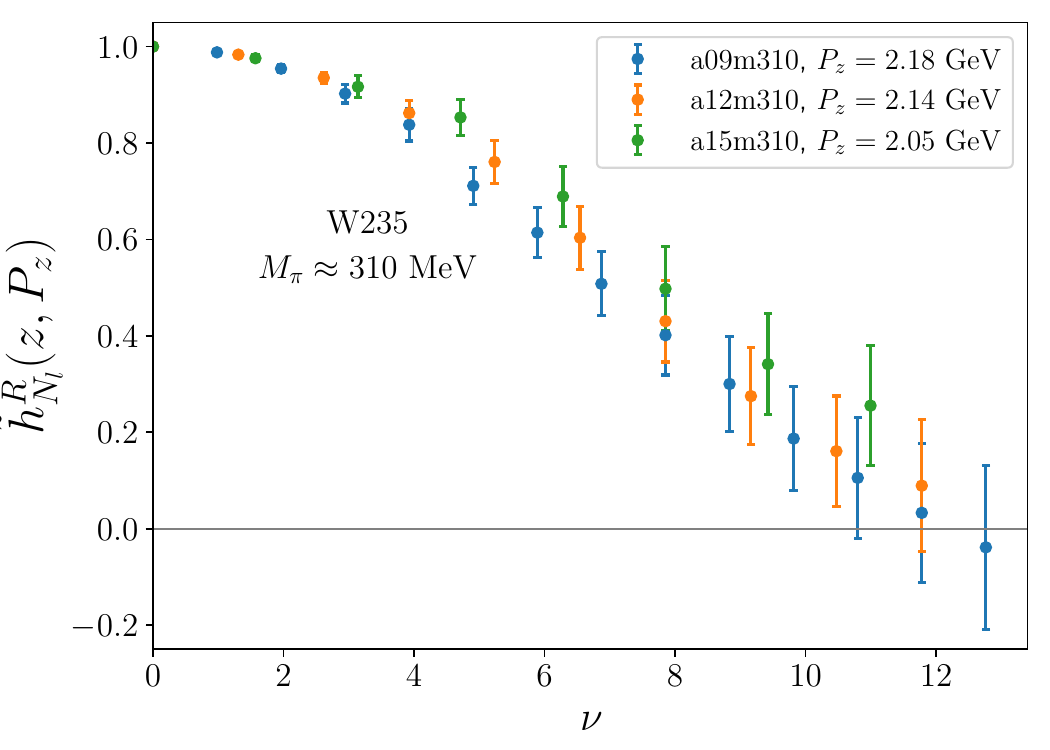}
\includegraphics[width=0.45\textwidth]{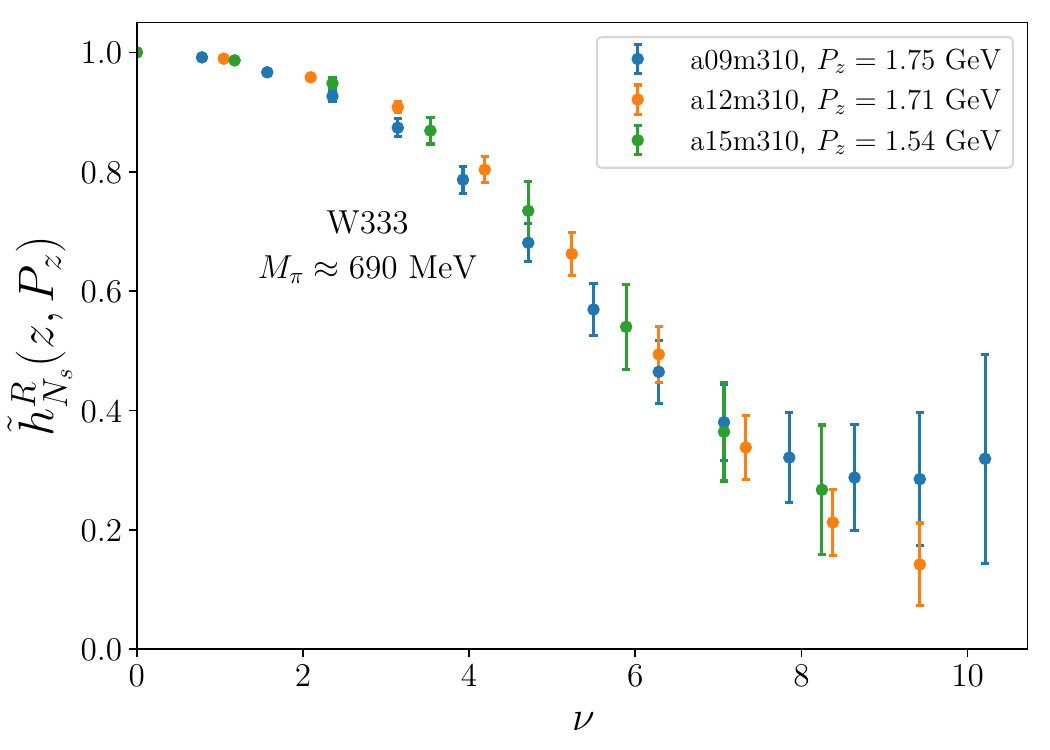}
\includegraphics[width=0.45\textwidth]{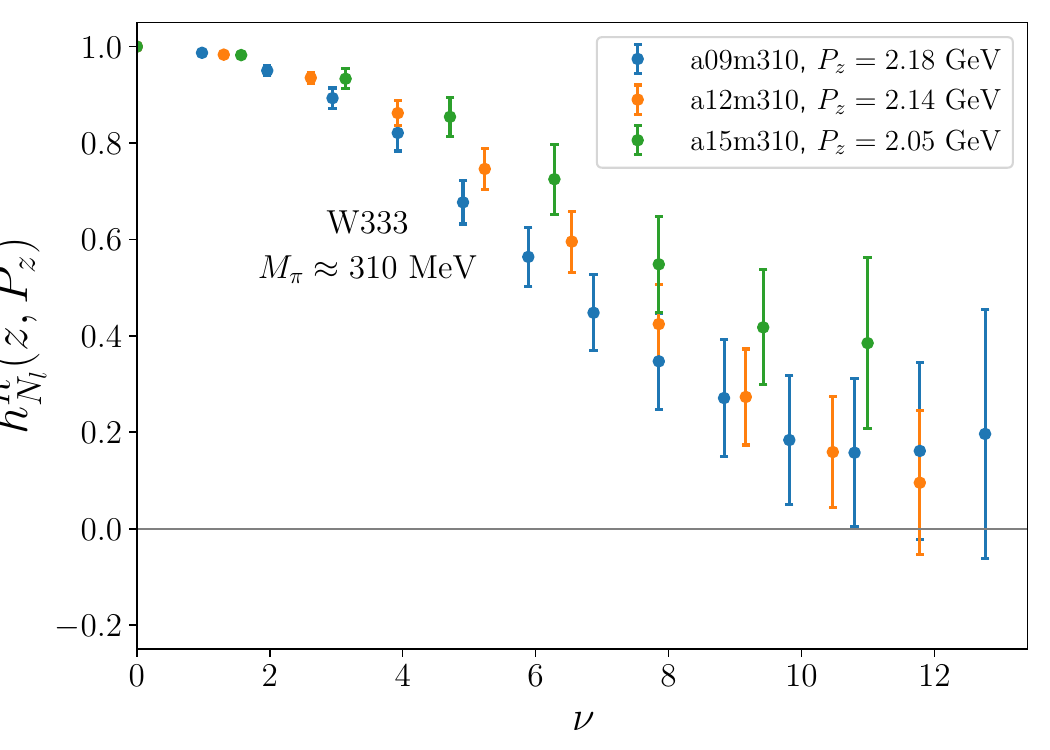}
\caption{
Renormalized matrix elements compared across each of the three lattice spacings $a \approx 0.09$, $0.12$, $0.15$ three smearing analyses at $M_\pi \approx 690$ (left column) and $310$ (right column) MeV for each of the three analyses W$123$ (top row), W$235$ (middle row), and W$333$ (bottom row).
\label{fig:renorm_MEs_ncls_ncll}
}
\end{figure*}

A similar comparison is seen for the $a \approx 0.12$~fm ensemble (middle panel in Fig.~\ref{fig:renorm_MEs_ncls}).
Here, W$235$ and W$333$ again share $\mathcal{T}_{\text{W}} = 3a^2$ but differ in whether the smearing is fixed in physical or relative units for renormalization, whereas W$123$ uses a distinct smearing level.
The central values remain closely aligned, though the uncertainties appear slightly reduced when the renormalization incorporates the additional statistical stabilizing effects observed at $a \approx 0.09$~fm.
The W$123$ results remain fully consistent with the other two analyses.
Finally, the $a \approx 0.15$~fm results (bottom panel in Fig.~\ref{fig:renorm_MEs_ncls}) show agreement among all three analyses within statistical uncertainties.
While the W$235$ smearing is smaller than that of W$333$ on this ensemble, the uncertainties for W$235$ are modestly reduced in several cases, which may again reflect the constraint introduced by the finer-lattice renormalization input.

A central purpose of the self-renormalization procedure is to suppress the dependence of the renormalized matrix elements on the lattice spacing.
In Fig.~\ref{fig:renorm_MEs_ncls_ncll}, we present the renormalized nucleon matrix elements for two representative momenta at both the heavier (left) and lighter (right) pion masses, evaluated on the three lattice spacings and for all three smearing analyses.
Across these results, the residual lattice-spacing effects are generally mild, with the data typically agreeing within one to two standard deviations.
The most noticeable deviations occur for the W$333$ ensemble at the lighter pion mass (bottom-right panel), where points around $\nu \approx 4$--$5$ lie slightly outside the two-standard-deviation band.
This may indicate a somewhat stronger interplay of smearing and discretization artifacts at lower pion mass, but such effects can be systematically addressed through a controlled continuum extrapolation.

Taken together, the three preceding figures suggest that, at the present statistical precision, the choice of smearing has only a modest impact on the extracted matrix elements.
More significantly, the self-renormalization approach appears robust for both fixed-physical and fixed-relative smearing conditions.
This is consistent with the expected scaling of the linear divergence, which is modulated by a factor proportional to $1/\sqrt{\mathcal{T}_\text{W}}$~\cite{Brambilla:2023vwm}.
For fixed relative flow time, one has $1/\sqrt{\mathcal{T}_\text{W}} \propto 1/a$, so that the corresponding linear divergence is effectively
absorbed into the leading term of Eq.~\ref{eq:renormalization-factor}.
Conversely, when the physical flow time is held fixed, the modulation can be accommodated through the renormalon term $m_0$, together with the nonperturbative function $g(z)$ in the parameterization.
We note that the self-renormalization framework was not originally developed with smeared operators in mind, and a more detailed investigation of this behavior will be pursued in future work.

To reconstruct the lightcone PDF, we first obtain the quasi-PDF in momentum space by performing a Fourier transformation of the renormalized spatial correlation function.
The quasi-PDF is then matched to the lightcone PDF using the appropriate perturbative matching kernel.
Since the Fourier transform is sensitive to the long-distance behavior of the matrix elements, a reliable description of the correlation function at separations larger than those directly accessible on the lattice is required to mitigate unphysical oscillatory artifacts in the quasi-PDF.

\begin{figure*}
\centering
\includegraphics[width=0.45\textwidth]{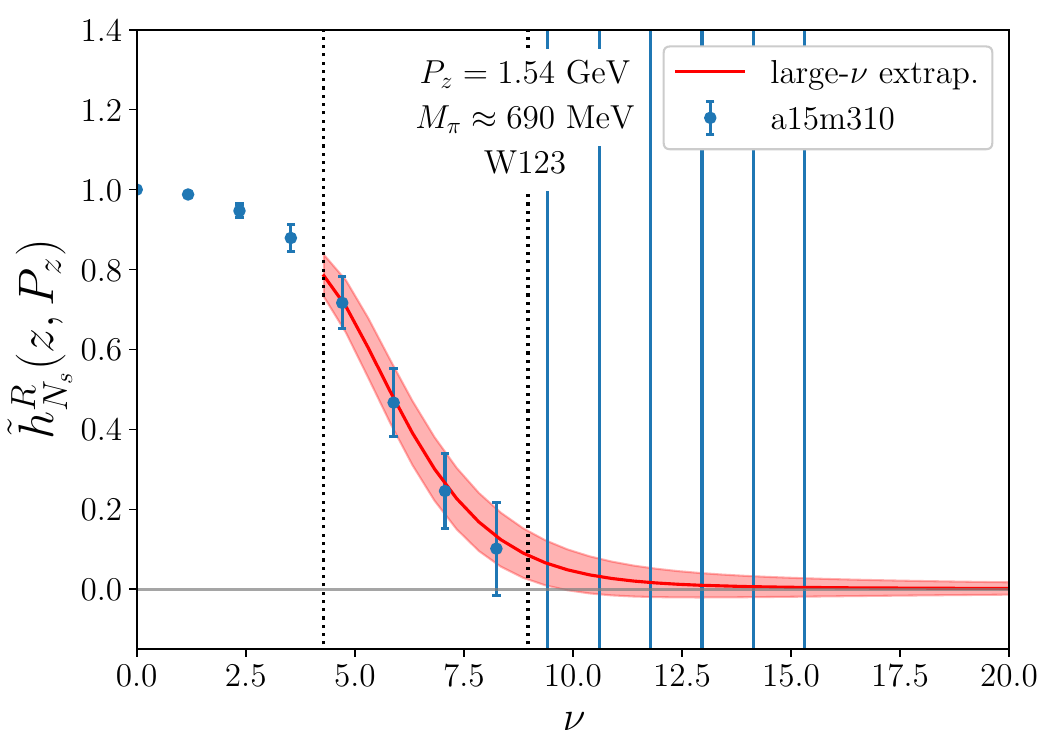}
\includegraphics[width=0.45\textwidth]{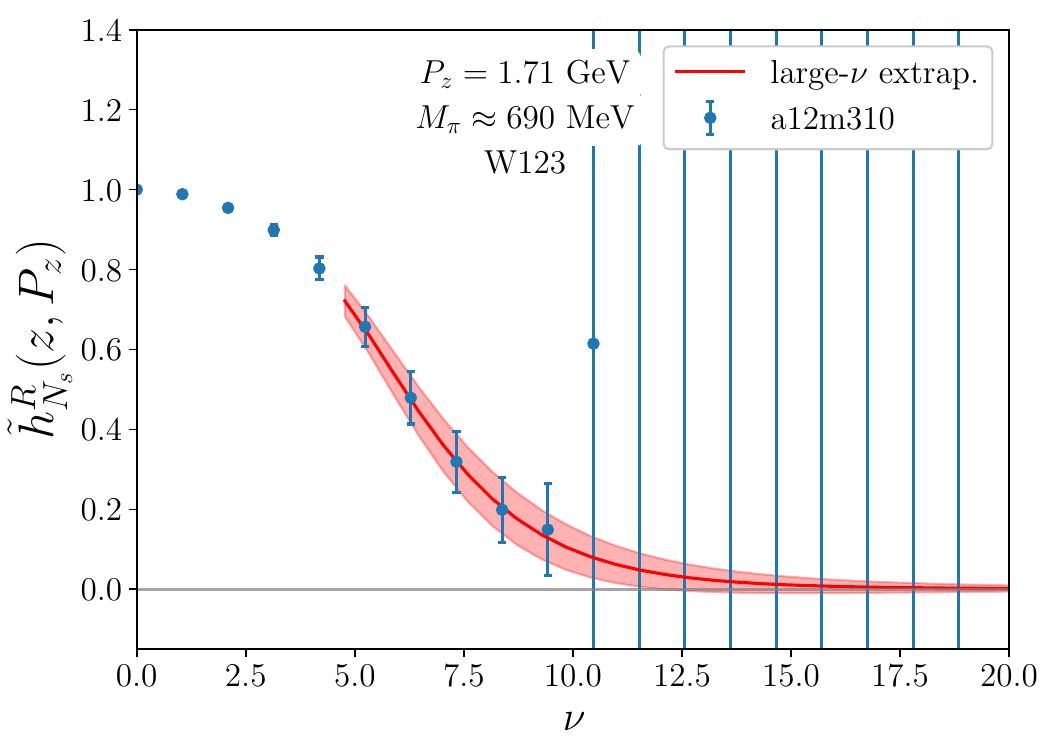}
\includegraphics[width=0.45\textwidth]{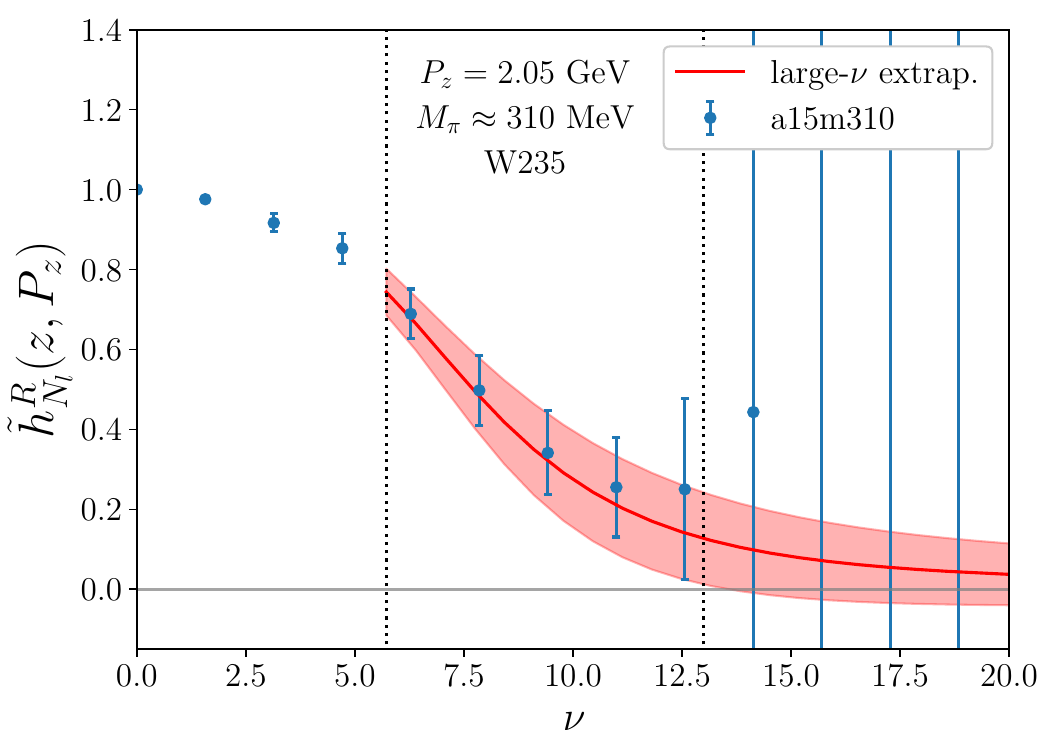}
\includegraphics[width=0.45\textwidth]{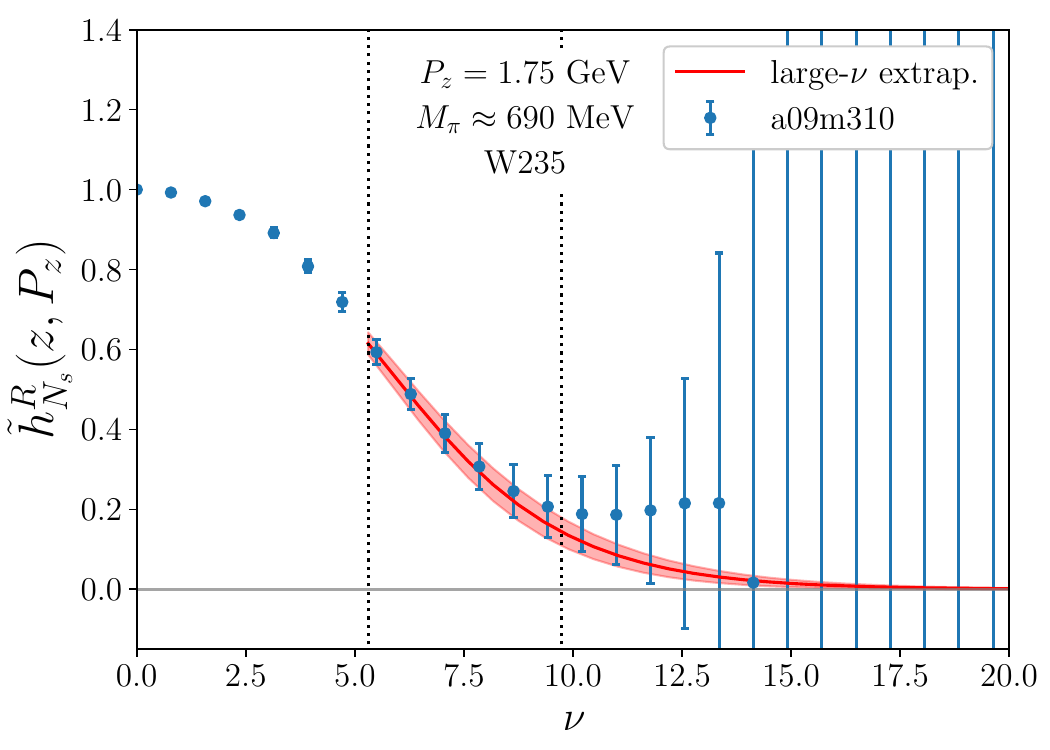}
\includegraphics[width=0.45\textwidth]{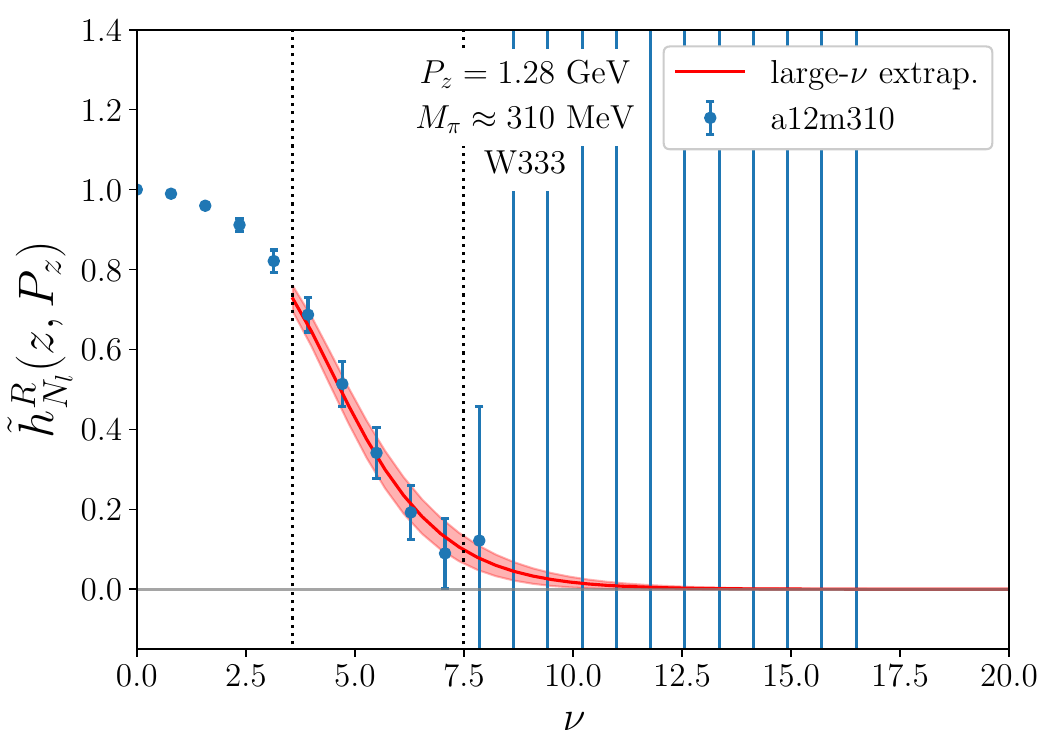}
\includegraphics[width=0.45\textwidth]{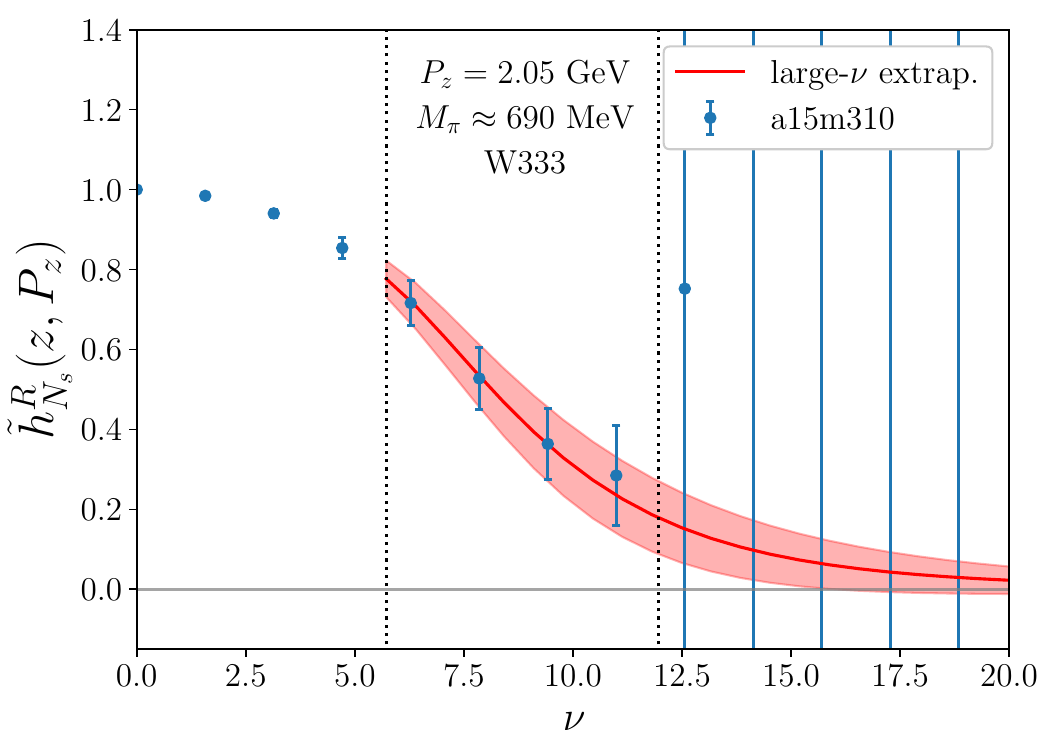}
\caption{
\label{fig:extrapolations}
Selected large-$\nu$ extrapolations at select momentum and ensemble for (top row) the W123 analysis, (middle row) the W235 analysis, and (bottom row) the W333 analysis.
We use data from $z \gtrsim 0.6$~fm across all examples and the black dashed lines illustrate the range of $z$ used in the fit
}
\end{figure*}

Following Ref.~\cite{Ji:2020brr}, we adopt the extrapolation form
\begin{equation}\label{eq:large-nu-form}
    h^{\text{R}}(z,P_z) \approx A\,\frac{e^{-m\nu}}{|\nu|^{d}},
\end{equation}
where $\nu = z P_z$ and $A$, $m$, and $d$ are fit parameters.
This functional form is motivated by theoretical considerations discussed in Ref.~\cite{Gao:2021dbh}.
Unlike the approach taken in our earlier work~\cite{Good:2025daz}, we now perform fully correlated fits of the parameters, which leads to improved control over statistical uncertainties.
Representative fits are shown in Fig.~\ref{fig:extrapolations}, with dashed lines indicating the region of data included in the fitting window.
In all cases, we exclude points with $z \lesssim 0.6$~fm from the extrapolation procedure, as this distance appears sufficient for the onset of the expected exponential decay behavior.
We include as many data points with reliable uncertainties as possible to ensure meaningful constraints on the fit parameters.
As intended, the extrapolated form significantly constrains the large-$|z|$ tail of the matrix elements compared to lattice data alone.

We note that the treatment of the long-distance behavior and the associated systematic uncertainties remains a subject of active study~\cite{Dutrieux:2025jed, Chen:2025cxr, Dutrieux:2025axb}.
Nevertheless, all fits yield $\chi^2/\text{dof}$ values consistent with unity, indicating that the adopted form provides an adequate description of the tail region.
Moreover, the contribution of this extrapolated part of the matrix elements to the intermediate-$x$ region where LaMET predictions are most reliable is expected to be modest.
A more comprehensive investigation of model and fit-range systematics will be pursued in future work.

\begin{figure}
\centering
\includegraphics[width=0.45\textwidth]{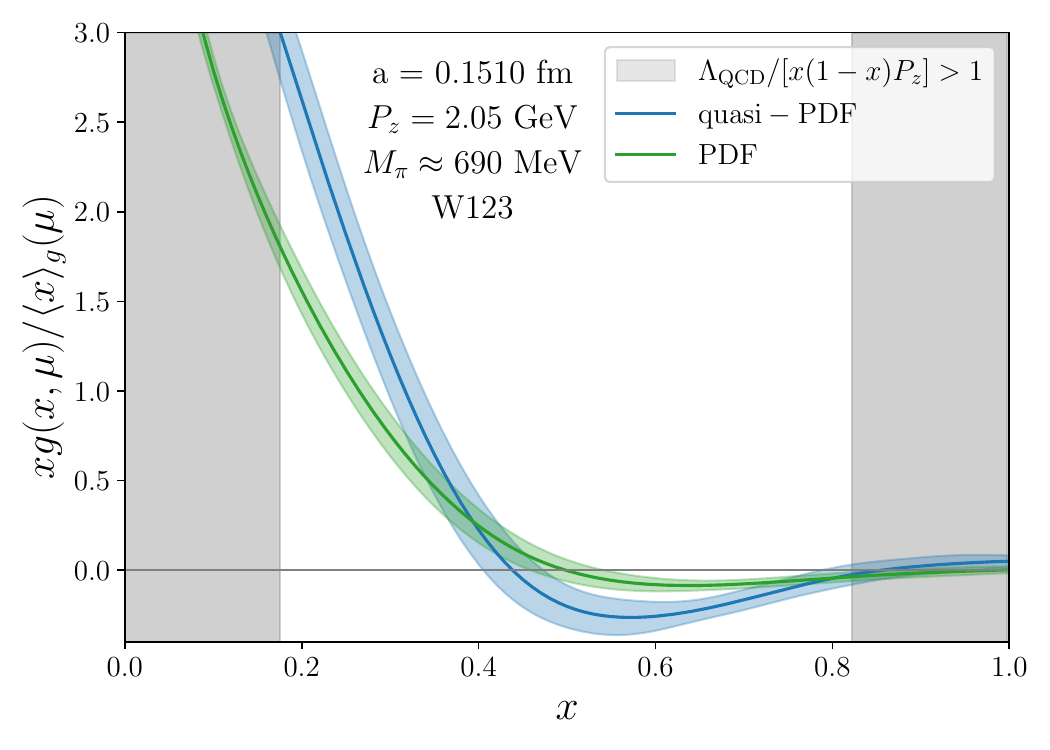}
\includegraphics[width=0.45\textwidth]{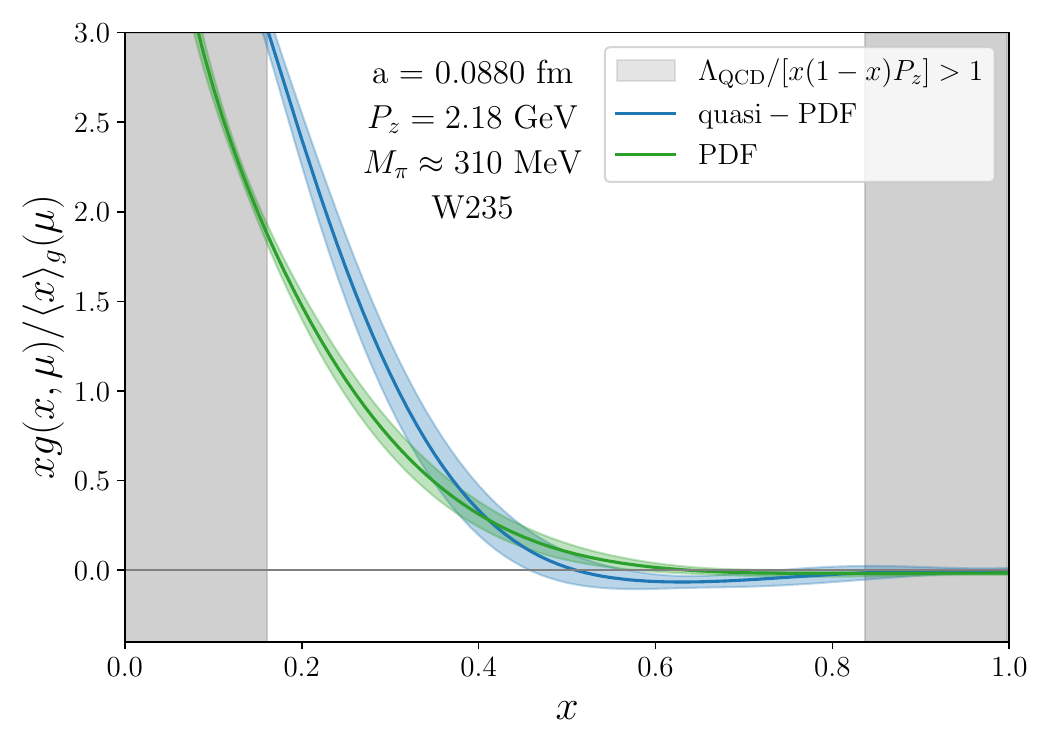}
\includegraphics[width=0.45\textwidth]{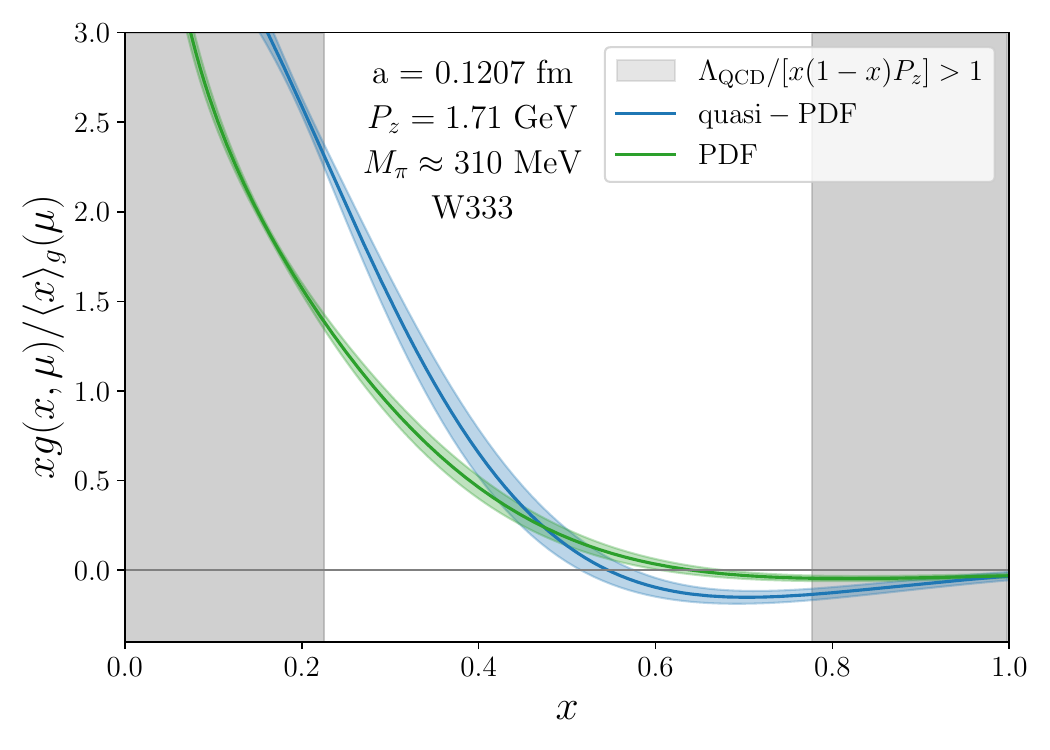}
\caption{
\label{fig:quasi-PDF_to_PDF}
A selection of quasi-PDFs and their corresponding lightcone PDFs after matching.
From top to bottom, the lattice parameters correspond to
$a \approx 0.15,\; 0.09,\; 0.12$~fm,
$M_\pi \approx 690$, $310$~MeV,
$P_z = 2.05,\; 2.18,\; 1.71$~GeV,
and smearing schemes W$123$, W$235$, and W$333$, respectively.
}
\end{figure}

The quasi-PDF is obtained by Fourier transforming the continuous matrix elements, which are constructed by combining the large-$\nu$ extrapolation with interpolations of the short-distance data.
The resulting quasi-PDF is then matched to the corresponding light-cone PDF using the perturbative matching kernel of Ref.~\cite{Good:2025daz}.
Quark mixing effects are neglected in this analysis.
Representative results for both quasi-PDFs and their matched lightcone counterparts are shown in Fig.~\ref{fig:quasi-PDF_to_PDF}.
In line with earlier studies of gluonic distributions within the LaMET framework~\cite{Good:2025daz}, the quasi-PDFs typically display mild negative regions which are significantly reduced after applying the matching.
Nevertheless, some residual negativity at large~$x$ persists in the matched PDFs, likely reflecting remaining lattice systematic effects, which we analyze in the following section.

\begin{figure*}
\centering
\includegraphics[width=0.45\textwidth]{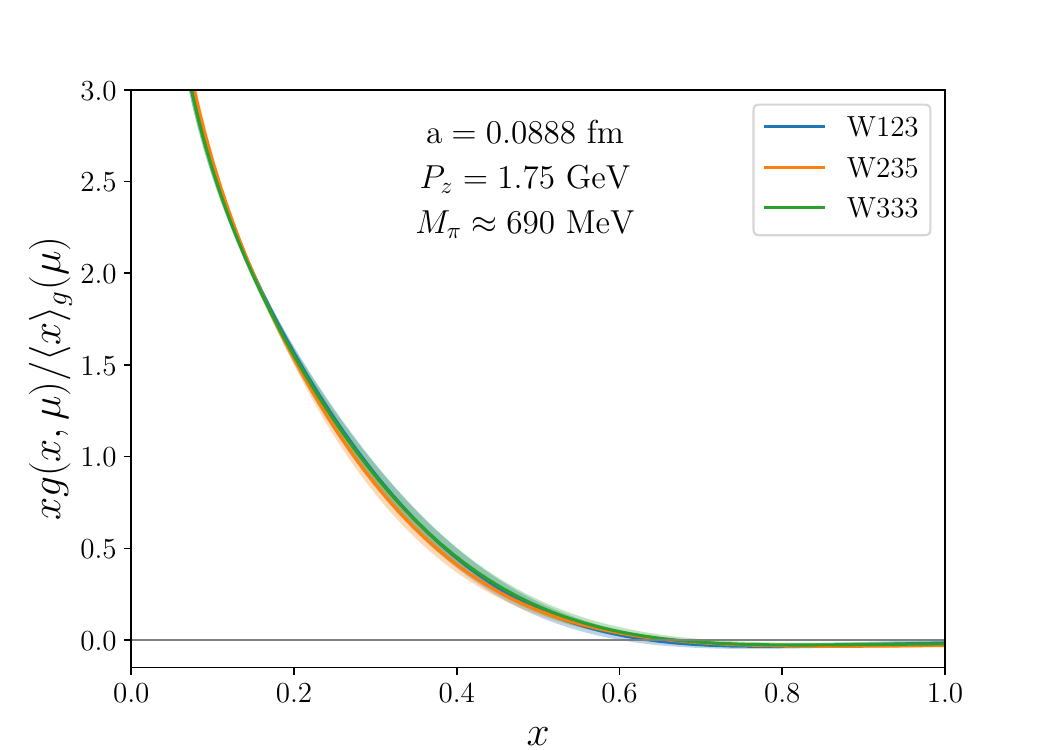}
\includegraphics[width=0.45\textwidth]{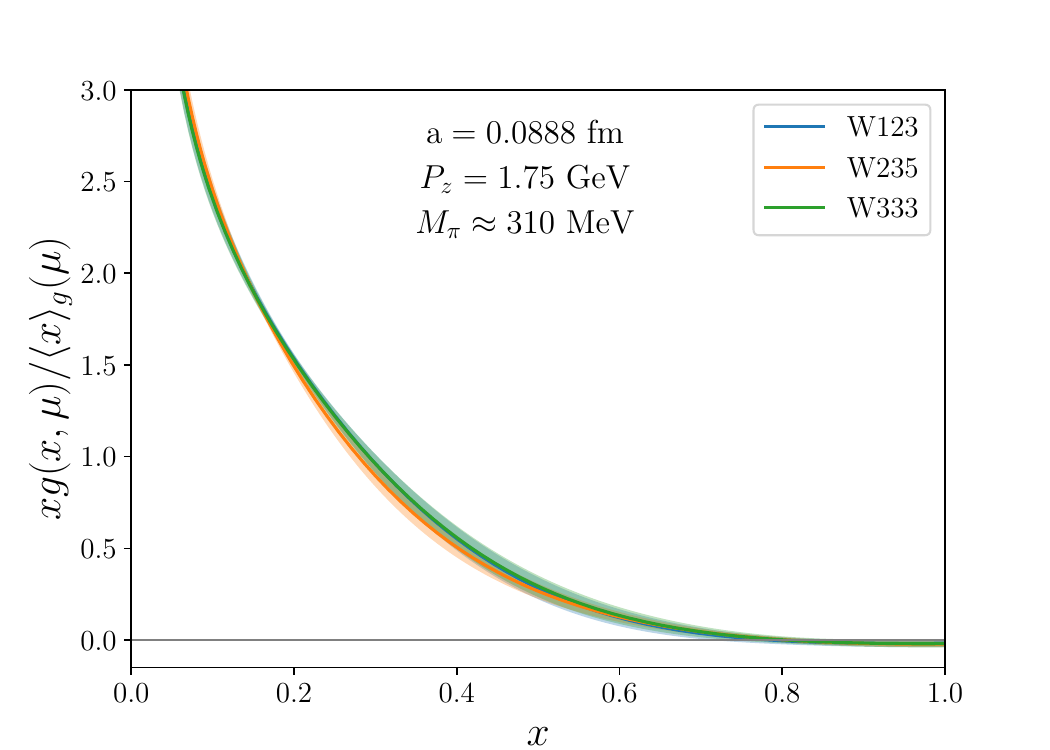}
\includegraphics[width=0.45\textwidth]{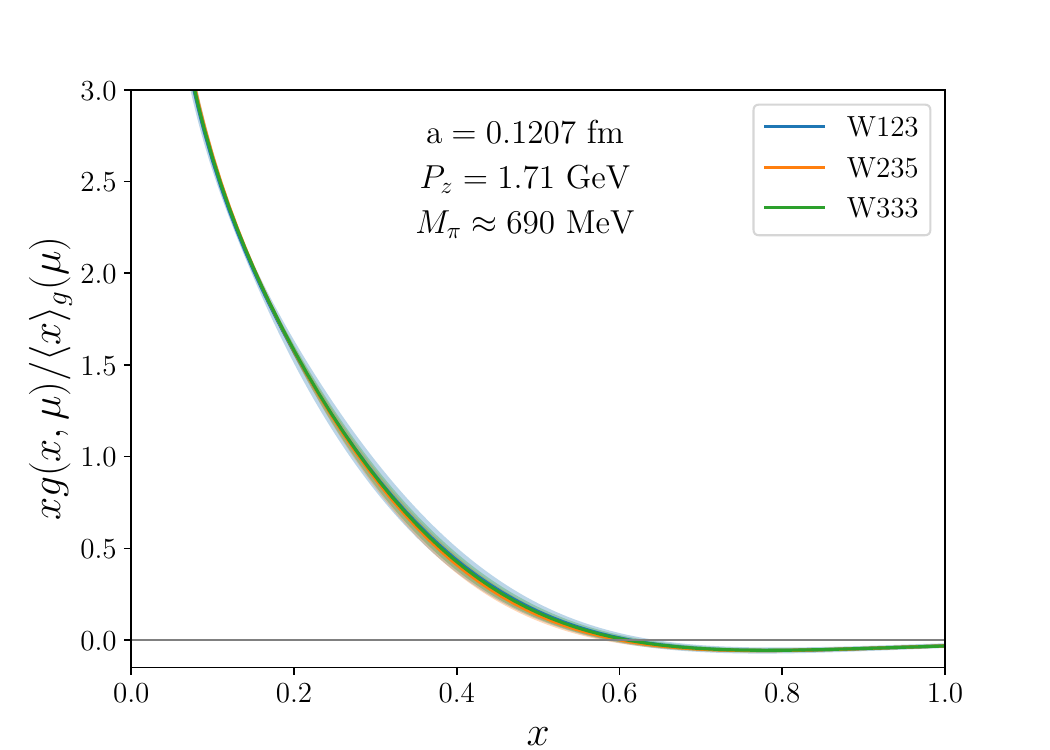}
\includegraphics[width=0.45\textwidth]{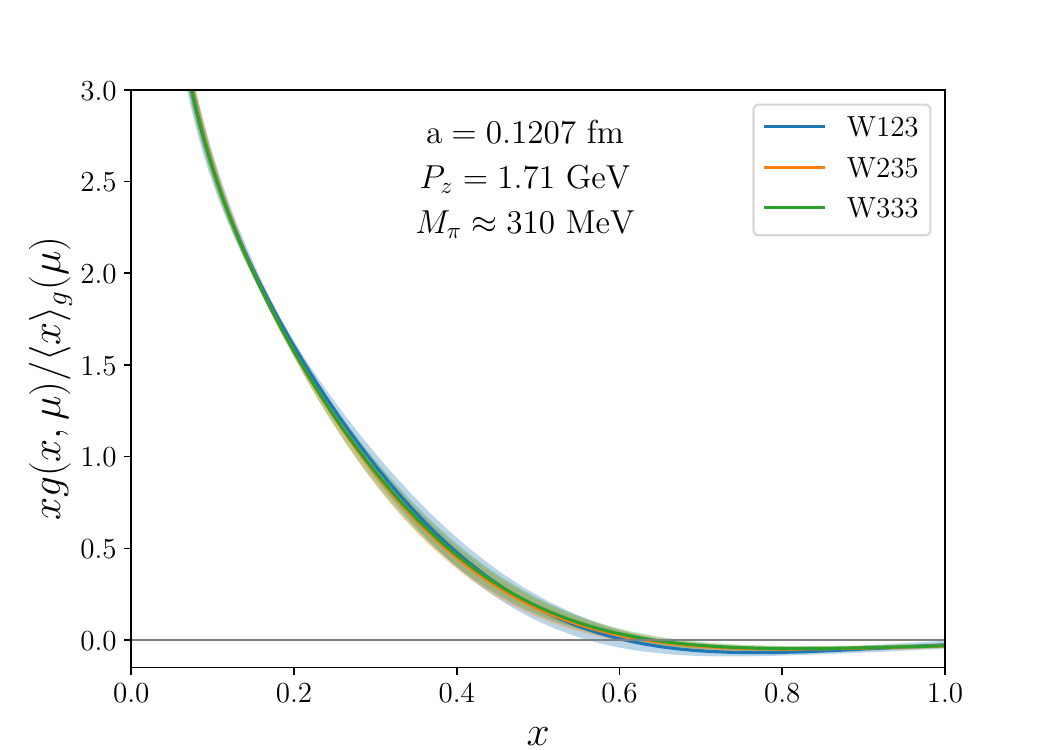}
\includegraphics[width=0.45\textwidth]{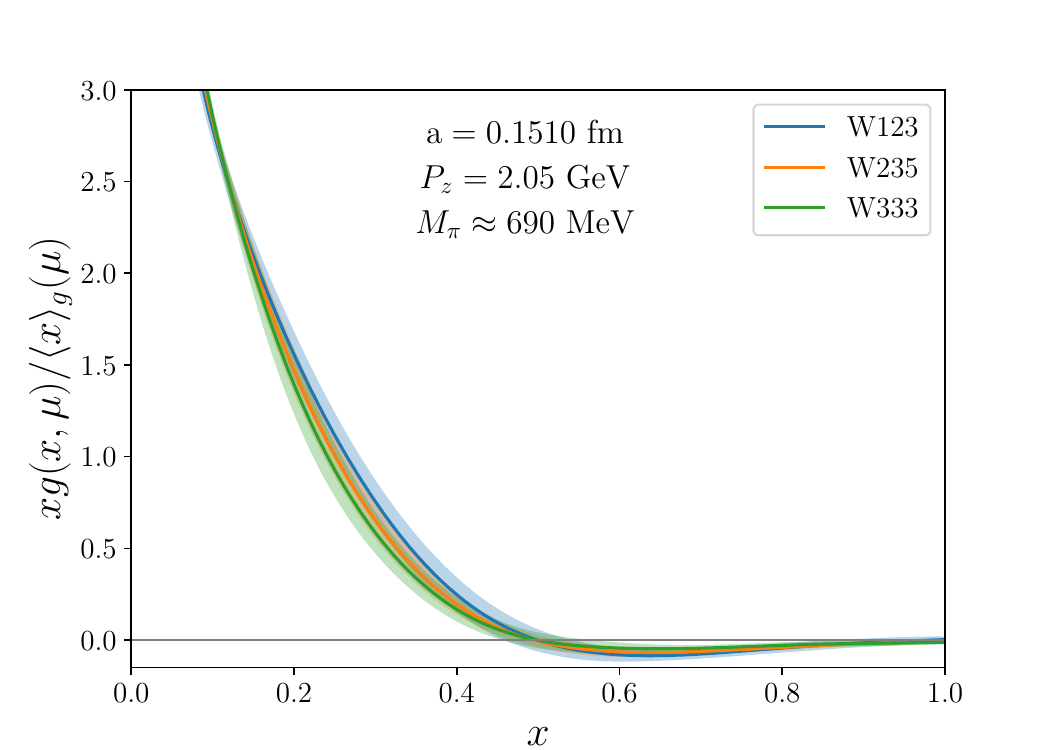}
\includegraphics[width=0.45\textwidth]{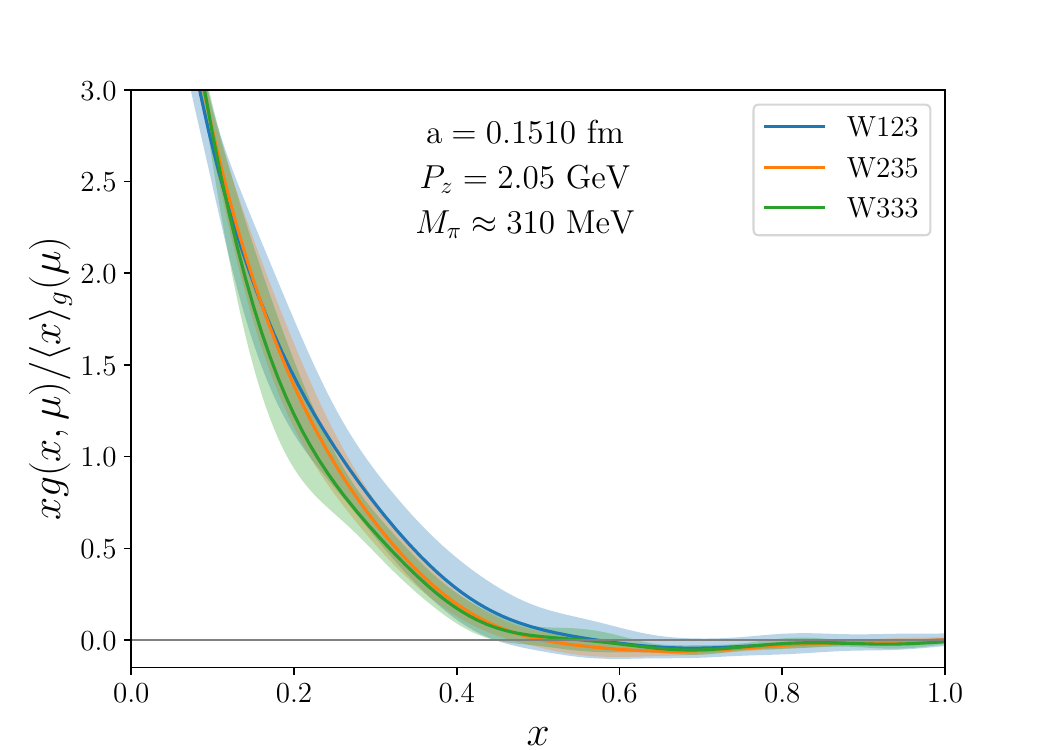}
\caption{
\label{fig:PDF_smear_dependency}
Lightcone PDFs from the three smearing analyses for select momenta across the three lattice spacings, $a\approx 0.09$ (top row), $0.12$ (middle row), and $0.15$ (bottom row) fm and two pion masses, $M_\pi \approx 690$ (left column) and $310$~MeV (right column).
}
\end{figure*}

\begin{figure*}
\centering
\includegraphics[width=0.45\textwidth]{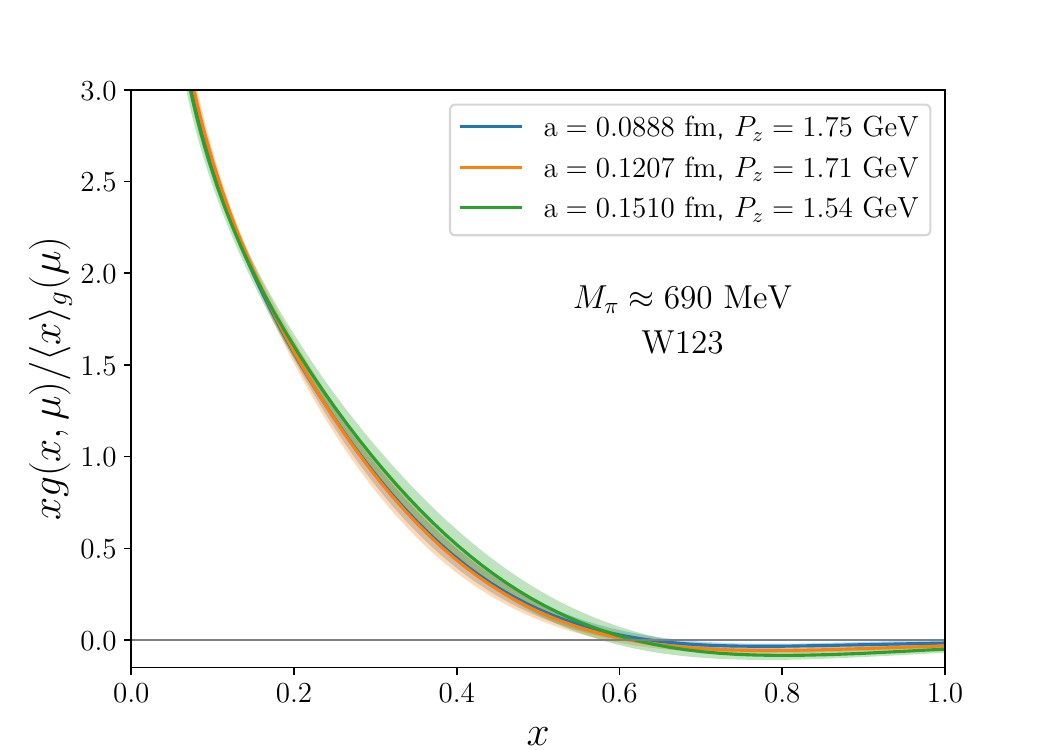}
\includegraphics[width=0.45\textwidth]{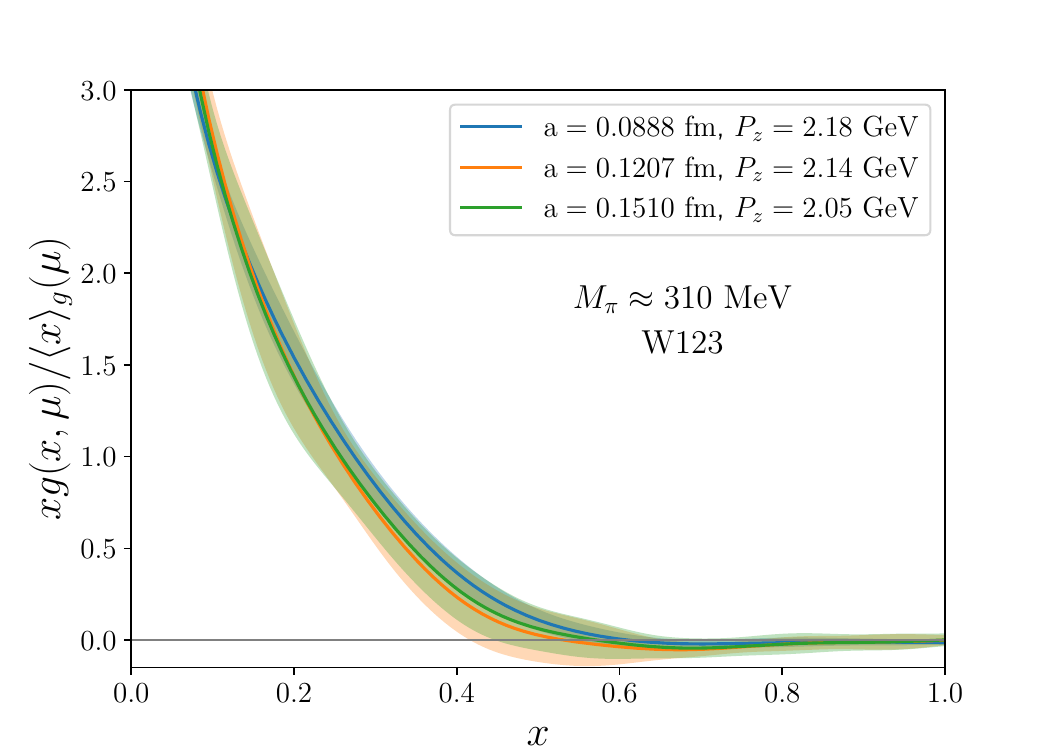}
\includegraphics[width=0.45\textwidth]{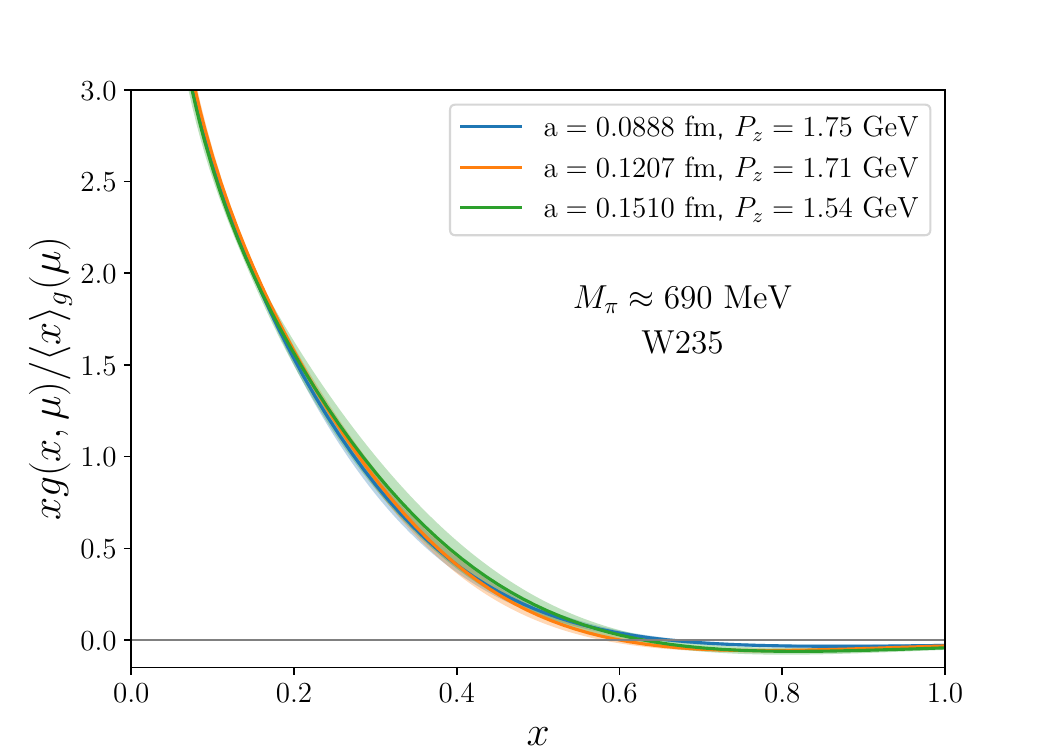}
\includegraphics[width=0.45\textwidth]{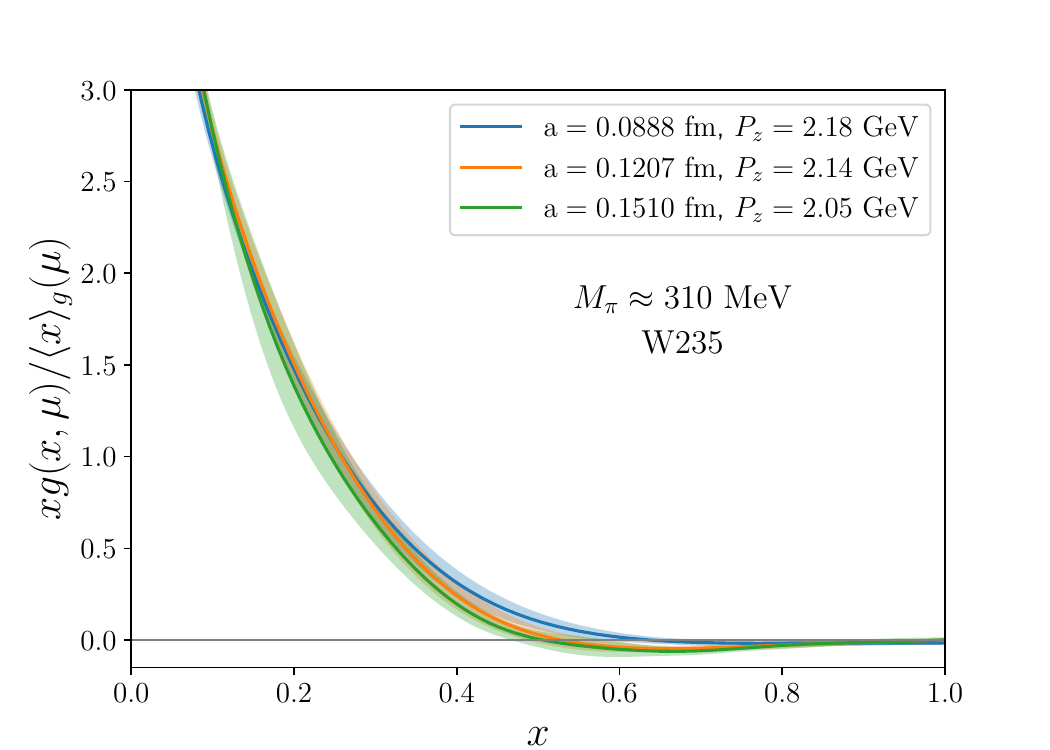}
\includegraphics[width=0.45\textwidth]{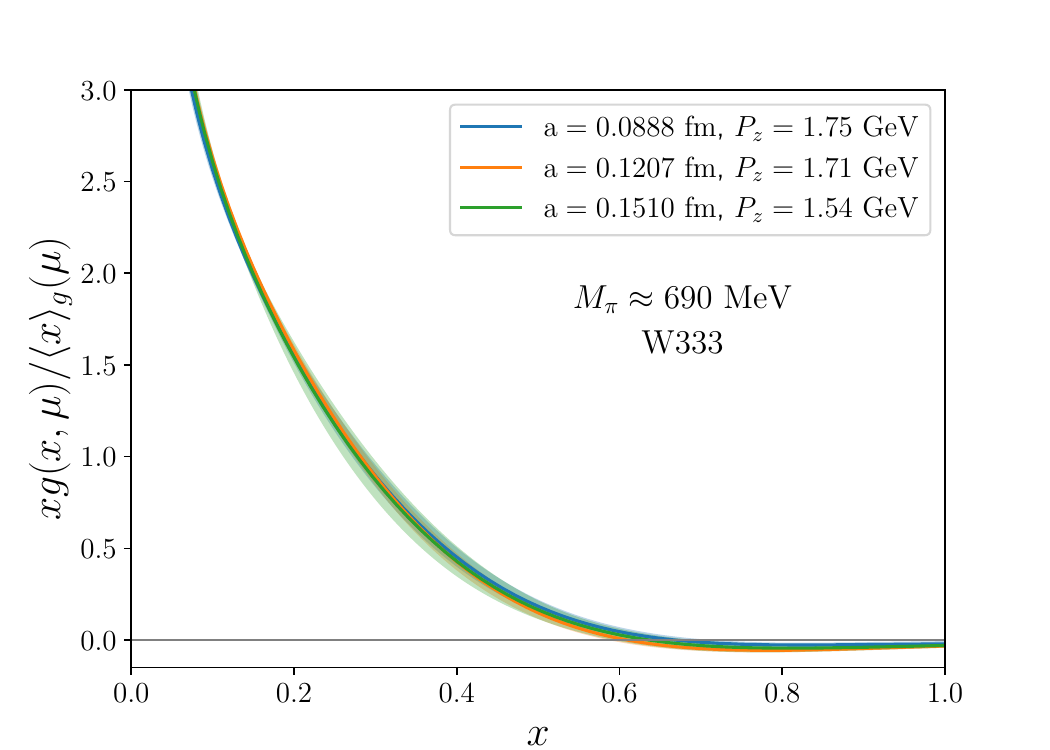}
\includegraphics[width=0.45\textwidth]{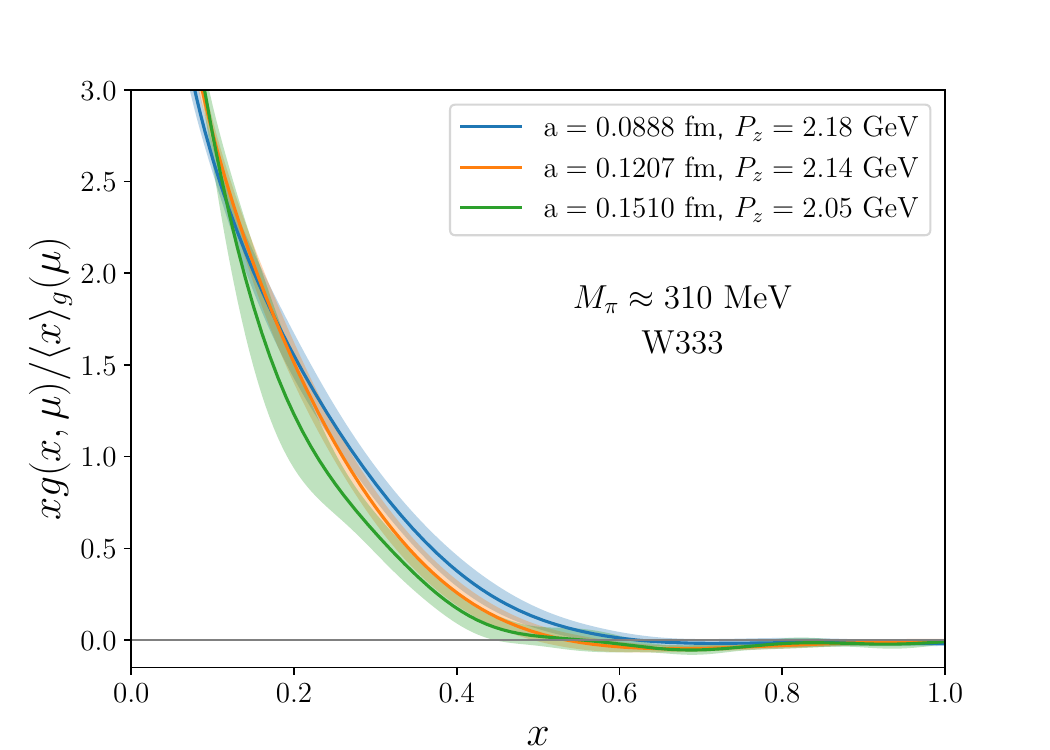}
\caption{
\label{fig:PDF_lattice_spacing_dependency}
Lightcone PDFs from the three smearing analyses W$123$ (top row), W$235$ (middle row), and W$333$ (bottom row) for select momenta across the three lattice spacings and two pion masses, $M_\pi \approx 690$ (left column) and $310$~MeV (right column).
}
\end{figure*}

With the extensive dataset available across multiple analyses, we can now explore the impact of various lattice systematics.
Up to this point, our discussion has focused solely on statistical uncertainties, which are now relatively well controlled.
The following results are therefore interpreted more qualitatively, acknowledging that systematic effects related to scale dependence and extrapolation may exceed the quoted statistical errors by several times.
We begin by revisiting the smearing dependence to verify the trends observed at the level of the matrix elements.
Figure~\ref{fig:PDF_smear_dependency} compares PDFs from different smearing analyses for several samples of lattice spacing, pion mass, and momentum.
Overall, the PDFs are remarkably consistent across the different smearing schemes, with the three PDFs in each plot falling comfortably within $1\sigma$ from each other.
This further demonstrates that self-renormalization is applicable to both the fixed-relative and fixed-physical smearing schemes, and removes any associated smearing effects.
This suggests that taking the limit of zero flow time explicitly for the fixed-physical smearing scheme is not necessary;
thus, we may move on to the other common systematics.
In Fig.~\ref{fig:PDF_lattice_spacing_dependency}, we plot PDFs from fixed smearing scheme and pion mass, at nearly consistent momenta, across the three lattice spacings.
As anticipated from the comparisons at the matrix-element level, lattice-spacing dependence is milder at the heavier pion mass (left column).
Here, all lattice spacings yield results consistent within roughly $1\sigma$ across all smearing schemes.
In contrast, the lighter pion mass (right column) exhibits more pronounced lattice-spacing effects, as expected.
The results are consistent within $1$--$2\sigma$ across all lattice spacings; however, it is clear that a reduction in statistical error would likely increase the tension between results at different lattice spacings.
One pleasing feature of the lattice-spacing effects is that smaller lattice spacings seem to lead to less negativity in the PDFs.
This can be seen best in the top left and middle right figures.
Some residual lattice-spacing dependence can be expected from the self-renormalization procedure, so finer lattice spacings and continuum extrapolations will be important for further controlling the lattice spacing effects, and reducing the unphysical negativity in the PDFs.

\begin{figure}
\centering
\includegraphics[width=0.45\textwidth]{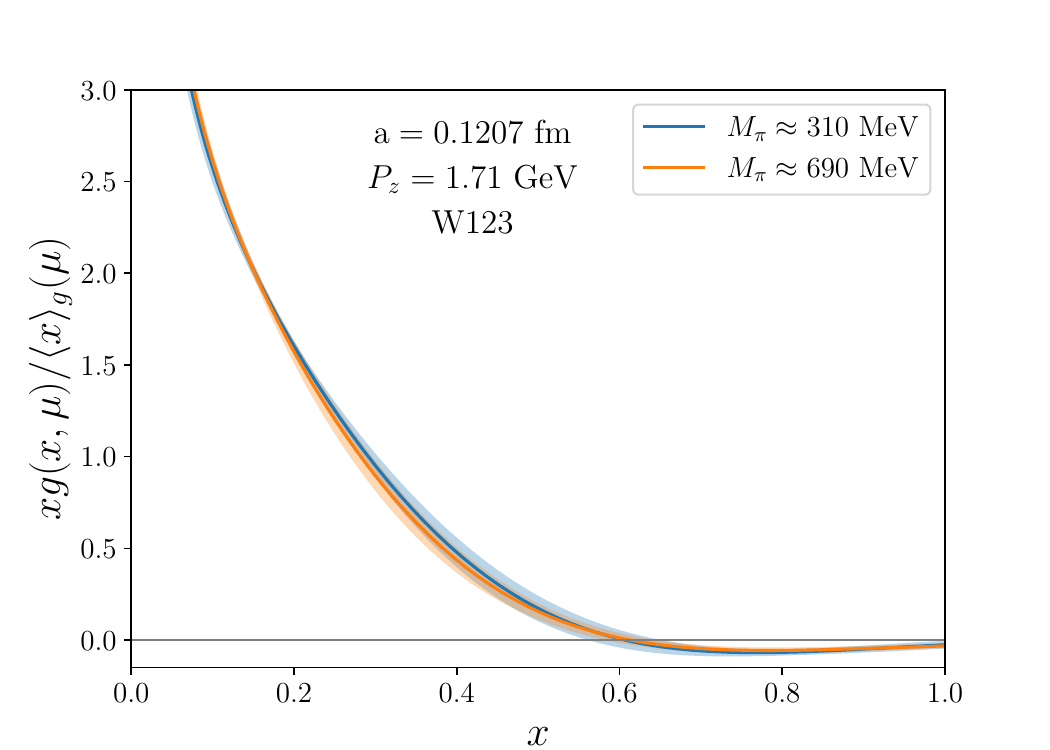}
\includegraphics[width=0.45\textwidth]{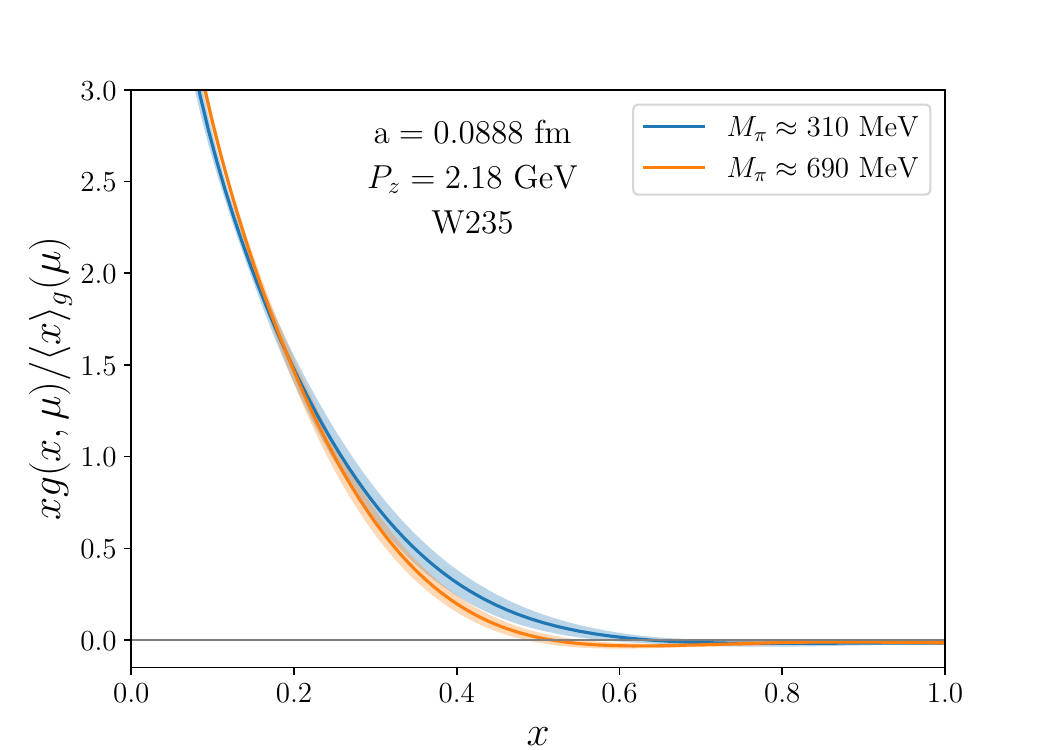}
\includegraphics[width=0.45\textwidth]{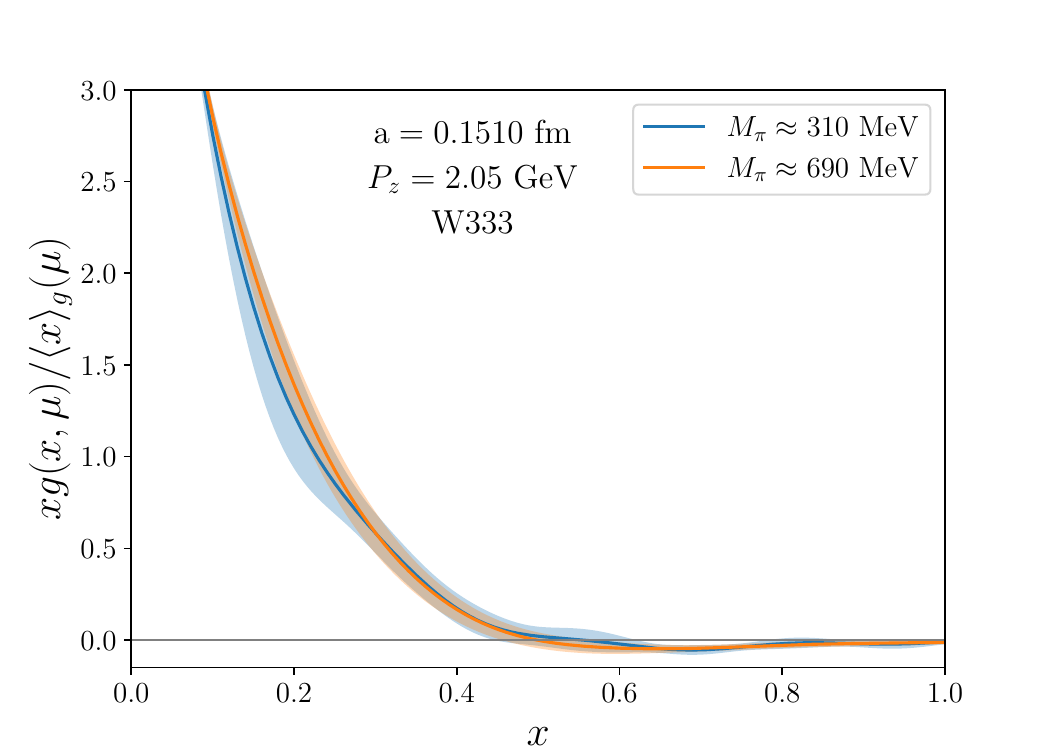}
\caption{
 \label{fig:PDF_pion_mass_dependency}
 Pion-mass dependence for a representative combination of smearing analyses, momentum, and lattice spacing.
 From top to bottom, the figures show $a\approx 0.12$, $0.09$, $0.15$~fm, $P_z \approx 1.71$, $2.18$, $2.05$~GeV, with W$123$, W$235$ and W$333$ smearing.
}
\end{figure}

\begin{figure}
\centering
\includegraphics[width=0.45\textwidth]{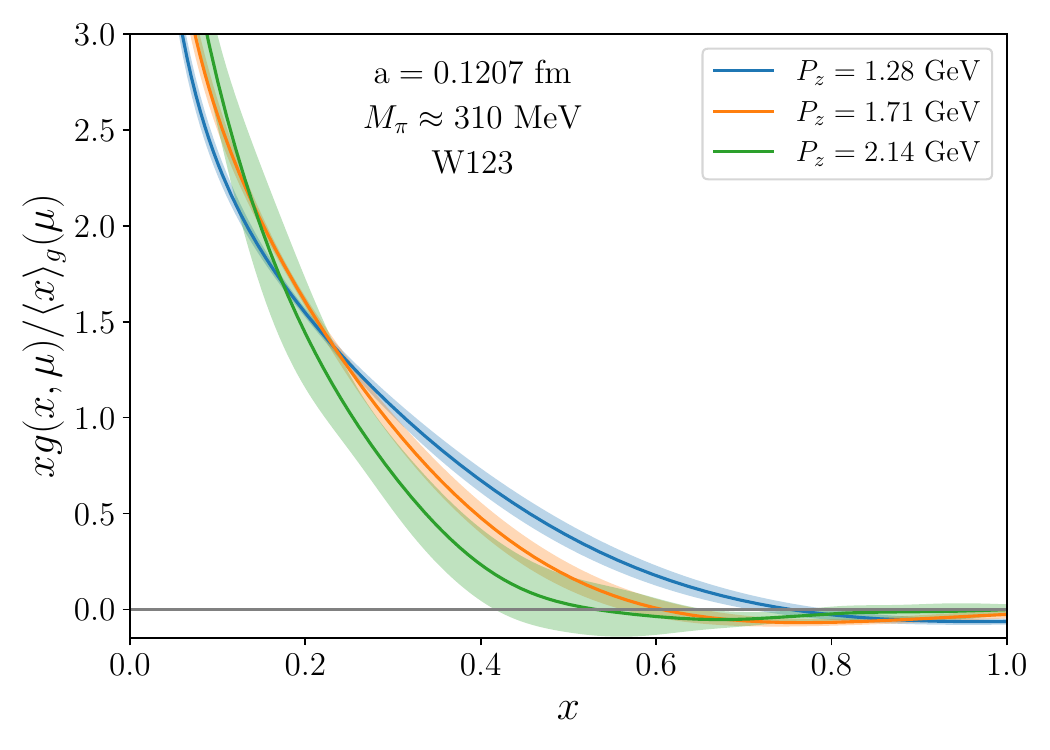}
\includegraphics[width=0.45\textwidth]{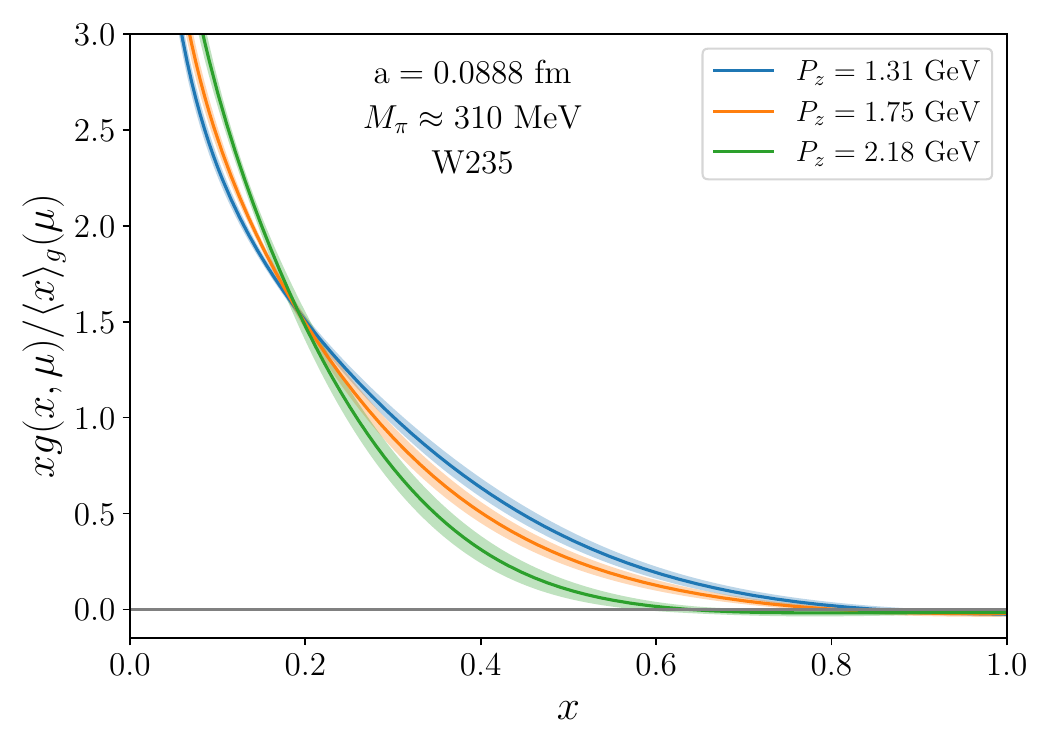}
\includegraphics[width=0.45\textwidth]{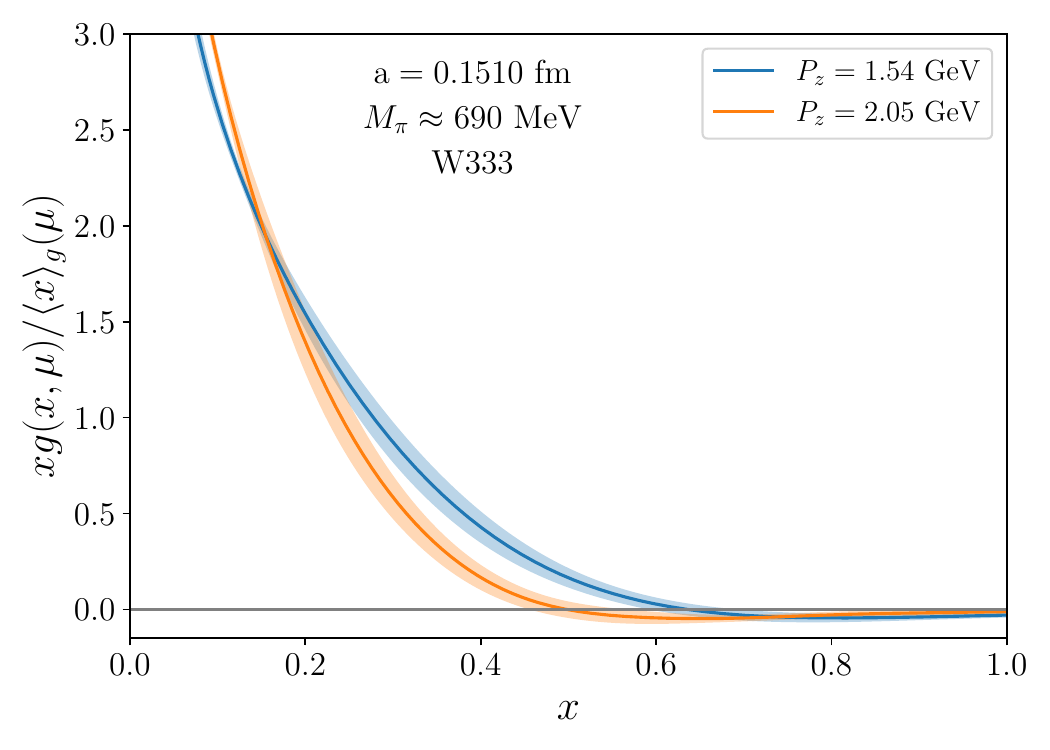}
\caption{
\label{fig:PDF_momentum_dependency}
Momentum dependence of the PDFs on an representative combination of lattice parameters.
From top to bottom, the figures show $a\approx 0.12$, $0.09$, $0.15$~fm, $M_\pi \approx 310$, $690$~MeV, with W$123$, W$235$ and W$333$ smearing.
}
\end{figure}

Now, we can move to the systematics which are not expected to be controlled by hybrid renormalization.
In Fig.~\ref{fig:PDF_pion_mass_dependency}, we show the pion-mass dependence of a sample of combinations of smearing analyses, momentum, and lattice spacing.
We see good consistency between the two pion masses, with a maximum of $2\sigma$ differences in the center plot.
This difference is not statistically significant, indicating that the pion-mass dependence is under control at the current level of precision. 
However, it would be valuable to confirm this in future work, either by performing a pion-mass extrapolation or by using improved algorithms and error reduction techniques that could make measurements at the physical pion mass more accessible. 
In particular, if the trend which is demonstrated most clearly in the center plot remains, physical pion mass may also contribute to a reduction in negativity of the PDF.
Finally, in Fig.~\ref{fig:PDF_momentum_dependency} we illustrate the momentum dependence of the PDFs for an arbitrary combination of lattice spacing, pion mass, and smearing choice;
these plots are representative of the many other combinations we have studied.
Here, strong momentum dependence is observed across all smearing schemes, with deviations exceeding $3\sigma$ in most regions.
This suggests that we may not have truly achieved the next-to-leading power accuracy in our LaMET expansion.
This can be remedied by leading-renormalon resummation~\cite{Zhang:2023bxs} and renormalization-group resummation~\cite{Su:2022fiu}, which have not yet been fully developed for the gluon.
Perhaps the deviation could be controlled by an infinite-momentum extrapolation; however, one must consider that the infinite-momentum result would only be accurate in the region $\Lambda_\text{QCD}/(x(1-x)P^\text{min}_z) < 1$.
We wish to further control the systematic effects and increase the momentum further before we apply infinite-momentum extrapolations.
We believe that pion mass and lattice-spacing effects are controllable via extrapolation at this point, but the momentum dependence is still quite large and needs to be understood further before an extrapolation is considered.

\section{Conclusions and Outlook}
\label{sec:conclusions}

In this work, we systematically investigated lattice artifacts arising from variations in gauge-link smearing, lattice spacing, pion mass, and nucleon boost momentum in the self-renormalized nucleon gluon PDF determination using the LaMET method.
We examined two smearing approaches, both using Wilson flow~\cite{Luscher:2010iy}: one with a fixed physical flow time and another with a fixed relative flow time on each ensemble.
The results show no significant differences in the renormalized matrix elements or in the extracted PDFs between the two methods.
Moreover, the dependence on the smearing level between different fixed physical flow times is negligible, suggesting that self-renormalization can control differences in smearing, and that extrapolation to zero flow time, is not even necessary.
The applicability of self-renormalization to Wilson-flowed data is expected, as the linear divergence should be modified by a term proportional to $1/\sqrt{\mathcal{T}_W} \propto 1/a$~\cite{Brambilla:2023vwm}.
In the case of fixed relative flow time, the self renormalization parameterization would be unchanged, while in the case of fixed physical flow time the modification of the linear divergence in fixed physical flow time can likely still be absorbed by the other terms in the self-renormalization parameterization.
We found very small pion-mass dependence and mild lattice spacing in the final PDFs, no more than about $2\sigma$.
These could be controlled fully via extrapolations to the continuum-physical limit and constrained further by calculations at smaller lattice spacings and pion masses.
On the other hand, we find strong momentum dependence, with differences corresponding to more than $3\sigma$ between the $P_z \approx 1.3$ and 2.2~GeV PDFs.
These effects may be caused by lack of leading-renormalon resummation (LRR)~\cite{Zhang:2023bxs} and renormalization-group resummation (RGR)~\cite{Su:2022fiu}, which have been developed for the flavor-nonsinglet parton distributions (see the recent review in Ref.~\cite{Lin:2025hka} for example results), but are not yet available for the gluon PDF.
These methods seem to improve momentum convergence in some of the leading twist-2 quark cases~\cite{Gao:2023ktu,Ding:2024saz}, by taking care of the leading $\Lambda_\text{QCD}/2xP_z$ corrections;
for our smallest-momentum PDFs which could be larger than $30\%$ at $x=0.5$~\cite{Zhang:2023bxs}.

Future studies of the gluon PDF, supported by increased computational resources, can further reduce both statistical and systematic uncertainties.
On the theoretical side, important questions remain regarding the optimal treatment of the Fourier transform in the quasi-PDF reconstruction.
Although higher boost momenta can in principle suppress leading LaMET systematics, achieving such boosts is difficult due to the rapidly deteriorating signal-to-noise ratio.
To enable higher-momentum calculations, one promising direction is the use of recently proposed kinematically enhanced two-point operators, which exhibit significantly improved signal-to-noise behavior compared with traditional nucleon operators~\cite{Zhang:2025hyo}.
Once the perturbative matching is worked out, we can do a guage fixed calculation of correlators without without Wilson lines~\cite{Gao:2023lny,Good:2024iur} to significantly improve the signal at long distances and yield better control over the large-$\nu$ extrapolation;
this may open the door to implementation of more sophisticated parametrizations of the extrapolation and exploration of Bayesian and machine-learning techniques~\cite{Chowdhury:2024ymm,Dutrieux:2024rem,Dutrieux:2025jed} to reduce model dependence.
Together with the finer lattice and lower pion masses, we can significantly improve the LaMET gluon PDFs from lattice QCD.

\section*{Acknowledgments}
AN and WG thank Yong Zhao, Yushan Su, Joshua Lin for their valuable suggestions and feedback on the earlier self-renormalized gluon PDF results.
We thank the MILC Collaboration for sharing the lattices used to perform this study.
The LQCD calculations were performed using the Chroma software suite~\cite{Edwards:2004sx}.
This research used resources of the National Energy Research Scientific Computing Center, a DOE Office of Science User Facility supported by the Office of Science of the U.S. Department of Energy under Contract No. DE-AC02-05CH11231 through ERCAP;
facilities of the USQCD Collaboration, which are funded by the Office of Science of the U.S. Department of Energy,
and supported in part by Michigan State University through computational resources provided by the Institute for Cyber-Enabled Research (iCER).
The work of AN and WG is partially supported by U.S. Department of Energy, Office of Science, under grant DE-SC0024053 ``High Energy Physics Computing Traineeship for Lattice Gauge Theory''.
The work of WG and HL is partially supported by the US National Science Foundation under grant PHY~2209424 and 2514533.
FY is supported by the U.S. Department of Energy, Office of Science, Office of Nuclear Physics through Contract No. DE-SC0012704, and within the framework of Scientific Discovery through Advanced Computing (SciDAC) award Fundamental Nuclear Physics at the Exascale and Beyond.

\bibliographystyle{unsrt}
\bibliography{main.bib}

\end{document}